\pdfoutput=1

\documentclass[aps,pra,10pt,tightenlines,longbibliography,notitlepage,amsmath,onecolumn,floatfix,superscriptaddress,eqsecnum]{revtex4-1}

\usepackage[dvipsnames]{xcolor}
\usepackage[colorlinks=true,bookmarks=false,linkcolor=NavyBlue,urlcolor=NavyBlue,citecolor=NavyBlue,breaklinks]{hyperref}
\usepackage{amsmath,amsfonts,amssymb,amsthm}
\usepackage{mathtools}
\usepackage{thmtools}
\renewcommand\thmcontinues[1]{\textbf{continued}}
\usepackage{tabularx}
\usepackage{graphicx}
\usepackage[italicdiff]{physics}
\usepackage{times}
\usepackage{enumitem}
\usepackage{setspace}

\newtheorem{theorem}{Theorem}
\newtheorem{corollary}{Corollary}
\newtheorem{proposition}{Proposition}[section]
\newtheorem{lemma}{Lemma}[section]
\newtheorem{example}{Example}

\newcommand{\new}[1]{{#1}}

\newcommand{\avg}[1]{\langle#1\rangle}

\newcommand{\Avg}[1]{\left\langle#1\right\rangle}

\newcommand{\bk}[1]{\left(#1\right)}
\newcommand{\Bk}[1]{\left[#1\right]}
\newcommand{\BK}[1]{\left\{#1\right\}}

\newcommand{\mc}[1]{\mathcal #1}
\newcommand{\mb}[1]{\mathbb #1}

\newcommand{\mr}[1]{\mathrm{#1}}
\newcommand{\Hc}[1]{\mathrm{H.c.}}

\newcommand{\Con}{\mathrm{Con}}

\newcommand{\Hel}{\mr{Hel}}
\newcommand{\KMB}{\mr{KMB}}

\renewcommand{\exp}{\operatorname{exp}}

\newcommand{\cgraphic}[2]{\centerline{\includegraphics[width=#1\textwidth]{#2}}}
\newcommand{\fig}[3]{
\begin{figure}[htbp!]
\cgraphic{#1}{#2}
\caption{\label{#2}#3}
\end{figure}
}
\begin{document}

\title{Quantum Onsager relations}

\author{Mankei Tsang}
\email{mankei@nus.edu.sg}
\homepage{https://blog.nus.edu.sg/mankei/}
\affiliation{Department of Electrical and Computer Engineering,
  National University of Singapore, 4 Engineering Drive 3, Singapore
  117583}

\affiliation{Department of Physics, National University of Singapore,
  2 Science Drive 3, Singapore 117551}

\date{\today}


\begin{abstract}
  Using quantum information geometry, I derive quantum generalizations
  of the Onsager rate equations, which model the dynamics of an open
  system near a steady state.  The generalized equations hold for a
  flexible definition of the forces as well as a large class of
  statistical divergence measures and quantum-Fisher-information
  metrics beyond the conventional definition of entropy production. I
  also derive quantum Onsager-Casimir relations for the transport
  tensors by proposing a general concept of time reversal and detailed
  balance for open quantum systems.  The results establish a
  remarkable connection between statistical mechanics and parameter
  estimation theory.
\end{abstract}

\maketitle

\section{Introduction}

In statistical mechanics, one is often interested in the dynamics of a
system close to a steady state. A crown jewel of near-equilibrium
statistical mechanics is the Onsager relations, which refer to a
symmetry of the transport coefficients for a broad class of
irreversible processes in diverse areas of physics and chemistry
\cite{onsager31,miller60,landau_stat,reichl}.

The basic idea of Onsager's theory is as follows. It was noticed that,
for a wide range of open systems near equilibrium, the rate of entropy
production, denoted as $R$, can be expressed as
\begin{align}
R &= f^j \dot y_j,
\label{R_intro}
\end{align}
where Einstein summation is assumed, $\{f^1, f^2,\dots\}$ are
thermodynamic forces, such as inverse temperatures and chemical
potentials at various parts of a system, each $y_j$ is the
thermodynamic variable conjugate to the force $f^j$, such as energy or
particle number, and $\dot y_j$ is the rate of change of $y_j$ in time
called a current. Moreover, the currents were often observed to follow
the relation
\begin{align}
\dot y_j &= O_{jk} f^k,
\label{rate_intro}
\end{align}
where $O$ is called the Onsager transport tensor.  Onsager argued
that, if the system satisfies detailed balance, then the tensor is
symmetric, viz.,
\begin{align}
O_{jk} &= O_{kj}.
\label{onsager}
\end{align}
Eq.~(\ref{onsager}) is called the Onsager relation.  The most famous
example may be the relation between the Peltier and Seebeck
thermoelectric coefficients \cite{miller60}.  Quantum versions of
Eq.~(\ref{onsager}) have been proposed most notably by Alicki
\cite{alicki76}, Spohn and Lebowitz \cite{spohn78}, and Lendi
\cite{lendi86}; see also
Refs.~\cite{lendi01,rodriguez20,salazar20,pusuluk21}.

Eq.~(\ref{onsager}) can also be generalized for variables that
transform nontrivially under time reversal or when time-reversal
symmetry is broken by, for example, an external magnetic field. The
resulting relations are called Onsager-Casimir relations
\cite{onsager31b,casimir45}, which relate the transport tensors before
and after the external field is reversed.

Most classical derivations of the Onsager theory assume Gaussian
statistics and linear dynamics, that is, $\dot y = L y$ for some rate
matrix $L$ \cite{landau_stat,reichl}.  The approach of Alicki
\cite{alicki76} and Lendi \cite{lendi86}, on the other hand, may be
the most general so far, as they are based on open quantum system
theory and do not assume Gaussian statistics or linear dynamics.  This
work further generalizes their approach using quantum information
geometry \cite{hayashi,amari}, showing that
Eqs.~(\ref{R_intro})--(\ref{onsager}) can hold for a flexible
definition of the forces beyond the thermodynamic setting and a large
class of statistical divergence measures, not just the conventional
entropy production. I show that the theory of Fisher information
\cite{hayashi,amari} is the hitherto unappreciated foundation
underlying Eqs.~(\ref{R_intro}) and (\ref{rate_intro}), thereby
establishing a remarkable connection between statistical mechanics and
parameter estimation theory.

To derive quantum Onsager-Casimir relations, one needs to invoke the
concept of time reversal. Time reversal for a closed system in terms
of an antiunitary operator is, of course, a well established concept
in quantum mechanics \cite[Sec.~4.4]{sakurai}, but its definition for
open quantum systems is far trickier.  Here I propose a definition
that generalizes Agarwal's approach \cite{agarwal73} and use it to
derive quantum Onsager-Casimir relations.

This work is organized as follows. Sec.~\ref{sec_method} introduces
the necessary background of open quantum system theory
\cite{alicki_lendi} and quantum information geometry
\cite{amari,hayashi}.  Sec.~\ref{sec_onsager} presents two key results
of this work: Theorem~\ref{thm}, which is a somewhat abstract analog
of Eqs.~(\ref{R_intro})--(\ref{onsager}) in the context of quantum
information geometry, and Theorem~\ref{thm2}, which uses
Theorem~\ref{thm} to prove a more physical generalization of
Eqs.~(\ref{R_intro}) and (\ref{rate_intro}). Sec.~\ref{sec_casimir}
proposes a general theory of time reversal for open quantum systems
and uses it to derive quantum Onsager-Casimir relations, generalizing
Eq.~(\ref{onsager}). Sec.~\ref{sec_discuss} discusses the significance
and potential applications of the results, while
Sec.~\ref{sec_conclusion} is the conclusion. The appendices contain
miscellaneous calculations and proofs that support the main text.

\section{\label{sec_method}Method}
\subsection{\label{sec_open}Open quantum system theory}
Let $\mc O(\mc H)$ be the set of all operators on a complex
Hilbert space $\mc H$ and
$\mc O_s(\mc H) \equiv \{A \in \mc O(\mc H): A = A^\dagger\}$
be the set of Hermitian operators, where $\dagger$ denotes the
Hermitian conjugate. Model the state of an open quantum system in the
Schr\"odinger picture at time $t \ge 0$ by a density operator
$\rho(t) \in \mc P(\mc H) \subset \mc O_s(\mc H)$, where $\mc P(\mc H)$
is the set of positive-definite operators with unit trace. Model the
dynamics by a quantum Markov semigroup
\begin{align}
\BK{\mc F(t) = \exp(\mc L t): t \ge 0},
\end{align}
where each $\mc F(t):\mc O(\mc H)\to \mc O(\mc H)$ is a completely
positive, trace-preserving (CPTP) map that maps Hermitian operators to
Hermitian operators and density operators to density operators.
$\mc L$ is called the semigroup generator and can be expressed in the
Gorini-Kossakowski-Sudarshan-Lindblad (GKSL) form
\cite{gorini76,lindblad76}.  Denoting the time derivative as
$\dot g(t) \equiv \pdv*{g(t)}{t}$, one can write
\begin{align}
\rho(t) &= \mc F(t) \rho(0),
&
\dot{\mc F}(t) &= \mc L \mc F(t),
&
\dot \rho(t) &= \mc L \rho(t).
\label{F}
\end{align}
In quantum optics, quantum thermodynamics, and quantum information
theory, quantum Markov semigroup is a basic tool for modeling the
effective dynamics of open quantum systems coupled with reservoirs
\cite{gardiner_zoller,wiseman_milburn,alicki_lendi,breuer,holevo19}. Its
use may be less prevalent in condensed matter physics \cite{weiss},
but its complete positivity and Markovianity make it the appropriate
basis for a quantum Onsager theory: complete positivity ensures
conformance with the second law of thermodynamics
\cite{lindblad75,alicki_lendi}, while Markovianity is a standard
assumption in the classical Onsager theory \cite{reichl,onsager53}.

Let $\sigma \in \mc P(\mc H)$ be a steady state of the semigroup,
viz.,
\begin{align}
\mc F(t) \sigma &= \sigma \quad\forall t,
&
\mc L \sigma &= 0.
\label{steady}
\end{align}
To model all the possible ways an experimenter can perturb the initial
state $\rho(0)$ away from the steady state, suppose that $\rho(0)$
depends on a $p$-dimensional parameter
$\theta \in \Phi \subseteq \mb R^p$. The dependence is modeled by a
function $\tau:\Phi \to \mc P(\mc H)$, so that
$\rho(0) = \tau(\theta)$. Suppose that $0 \in \Phi$ without loss of
generality and $\tau(0)$ coincides with the steady state, viz.,
$\tau(0) = \sigma$. To model the near-equilibrium condition in the
Onsager theory, assume that the initial state $\rho(0)$ is close to
the steady state, such that it can be approximated by
the first-order Taylor series 
\begin{align}
\rho(0) &= \tau(\epsilon v) = 
\sigma + \epsilon v^j \partial_j \tau + o(\epsilon),
\label{taylor0}
\end{align}
where
$\partial_j g(\theta) \equiv \pdv*{g(\theta)}{\theta^j}|_{\theta =
  0}$, $v \in \mathbb R^p$ is a vector, $\epsilon \in \mathbb R$ is an
infinitesimal real number, and $o(\epsilon)$ denotes terms
asymptotically smaller than $\epsilon$. \new{$\theta$ is replaced by
  $\epsilon v$ in the Taylor series to clarify the order of each term
  with respect to the infinitesimal quantity $\epsilon$.}  $\rho(t)$
can then be approximated as
\begin{align}
\rho(t) &= \mc F(t) \rho(0) = 
\sigma + \epsilon v^j \mc F(t) \partial_j \tau + o(\epsilon).
\label{taylor}
\end{align}
The following two physical examples will be used to illustrate the
theory throughout this work.

\begin{example}[label=exa_gibbs]
  Consider a system with a steady state given by a generalized Gibbs
  state \cite{jaynes57}
\begin{align}
\sigma &= \frac{\exp(\lambda^j G_j)}{\trace \exp(\lambda^jG_j)},
\label{gibbs_sigma}
\end{align}
where each $\lambda^j \in \mb R$ is a real parameter and each
$G_j \in \mc O_s(\mc H)$ is a Hermitian operator. In practice, $\{G_j\}$
may be a set of energy and particle-number operators for different
parts of the system. $\{\lambda^j\}$ are then proportional to the
inverse temperatures and chemical potentials of the subsystems at the
steady state. Suppose that the parameters can be controlled by a reservoir
coupled to the system. Some time before $t = 0$, some controls at the
reservoir are switched on to prepare the system in a perturbed Gibbs state
at $t = 0$ given by \cite[Eq.~(4.6.19)]{kubo}
\begin{align}
\rho(0) &= \tau(\epsilon v) = 
\frac{\exp[(\lambda^j+\epsilon v^j) G_j]}
{\trace \exp[(\lambda^j + \epsilon v^j) G_j]}.
\label{gibbs}
\end{align}
The controls are then switched off at $t = 0$, such that the dynamics
is modeled by the semigroup $\mc F(t)$ for $t\ge 0$, $\sigma$ is the
steady state of the dynamics, and the system evolves from the initial
state $\rho(0)$ that is slightly perturbed from $\sigma$.
\end{example}
\begin{example}[label=exa_kick]
  Suppose that the system is in a steady state $\sigma$ for $t < 0$
  but experiences a gentle kick right before $t = 0$, modeled by a
  unitary operator
\begin{align}
U = \exp(-i\epsilon v^j G_j),
\end{align}
where each $G_j \in \mc O_s(\mc H)$ is a generator of the unitary. The
kick in practice may be a force on a mechanical system, an applied
voltage on charges, a magnetic field on spins, a source generating
electromagnetic fields, a phase modulation on light, or any
combination of them on a hybrid system. Then the state at $t = 0$
becomes
\begin{align}
\rho(0) &= \tau(\epsilon v) = U \sigma U^\dagger = \sigma - i \epsilon
v^j\Bk{G_j,\sigma} + o(\epsilon).
\label{kick}
\end{align}
The dynamics for $t \ge 0$ is again assumed to be modeled by
$\mc F(t)$.
\end{example}

In the following, a physicists' level of mathematical rigor will be
adopted, in the sense that functions are as continuous as needed for
differentiation, terms denoted by $o(\epsilon^n)$ are always assumed
to be negligible, all proofs assume finite-dimensional spaces, and the
results will be applied to a bosonic field in Example~\ref{exa_kick}
despite the finite-dimensional assumptions in the proofs.

\subsection{\label{sec_info}Quantum information geometry}
To introduce information geometry to the problem, assume a divergence
measure between $\rho(t)$ and $\sigma$ that can be approximated as
\begin{align}
D(\rho(t)\Vert\sigma) &= 
\frac{1}{2}\epsilon^2 v^j J_{jk}(t)  v^k + o(\epsilon^2),
\label{D}
\\
J_{jk}(t) &\equiv 
\trace \Bk{\mc F(t)\partial_j\tau} \mc E_\sigma^{-1}\Bk{\mc F(t)\partial_k\tau},
\label{J}
\end{align}
where $\trace$ denotes the trace and
$\mc E_\sigma:\mc O(\mc H)\to\mc O(\mc H)$ is a linear map that
depends on $\sigma$. I call $\mc E_\sigma$ a density map.  To make $J$
well behaved, the density map is required to be self-adjoint and
positive-definite with respect to the Hilbert-Schmidt inner product
\begin{align}
\Avg{A,B} &\equiv \trace A^\dagger B
\quad
\forall A,B \in \mc O(\mc H).
\label{HS}
\end{align}
Then $J$ is positive-semidefinite and can be regarded as a Riemannian
metric. In estimation theory, $J$ is called a quantum Fisher
information \cite{amari,hayashi,petz}. Another common form of
Eq.~(\ref{J}) can be obtained by defining an operator
$X_j(t) \in \mc O(\mc H)$ as the solution to
\begin{align}
\partial_j\mc F(t)\tau &= \mc F(t)\partial_j\tau = \mc E_\sigma X_j(t),
\label{score}
\end{align}
such that 
\begin{align}
J_{jk}(t) &= \Avg{X_j(t),X_k(t)}_\sigma,
\label{J2}
\end{align}
where the weighted inner product $\avg{\cdot,\cdot}_\sigma$ is defined as
\begin{align}
\Avg{A,B}_\sigma &\equiv \Avg{A,\mc E_\sigma B} =\trace A^\dagger \mc E_\sigma B
\quad
\forall A,B \in \mc O(\mc H).
\label{weighted}
\end{align}
$X_j(t)$ is called a score or a logarithmic derivative in information
geometry \cite{amari,hayashi,petz,semi_prx}. In what follows, I assume
that $\mc E_\sigma$ maps Hermitian operators to Hermitian operators (a
property called symmetric), such that each score is Hermitian, each
$J_{jk}(t)$ entry is real, and the $J$ tensor is symmetric, viz.,
\begin{align}
J_{jk}(t) = J_{kj}(t).
\label{Jsym}
\end{align}
In information geometry, $v^j X_j(t)$ is \new{a representation} of the
tangent vector that points from the steady state $\sigma$ to $\rho(t)$
in the space of density operators \new{\cite{hayashi,amari}}, as
$v^jX_j(t)$ determines $\rho(t)$ via
$\rho(t) = \mc F(t) \tau(\epsilon v) = \sigma + \epsilon \mc
E_\sigma\Bk{v^j X_j(t)} + o(\epsilon)$ in the first order.  Given the
Fisher metric, the tangent vector can be assigned a length
$[v^j\avg{X_j(t),X_k(t)}_\sigma v^k]^{1/2} = [v^j J_{jk}(t)
v^k]^{1/2}$, and the length determines the divergence between
$\rho(t)$ and $\sigma$ via Eq.~(\ref{D}). Fig.~\ref{pushforward}
illustrates these geometric concepts.

\fig{0.6}{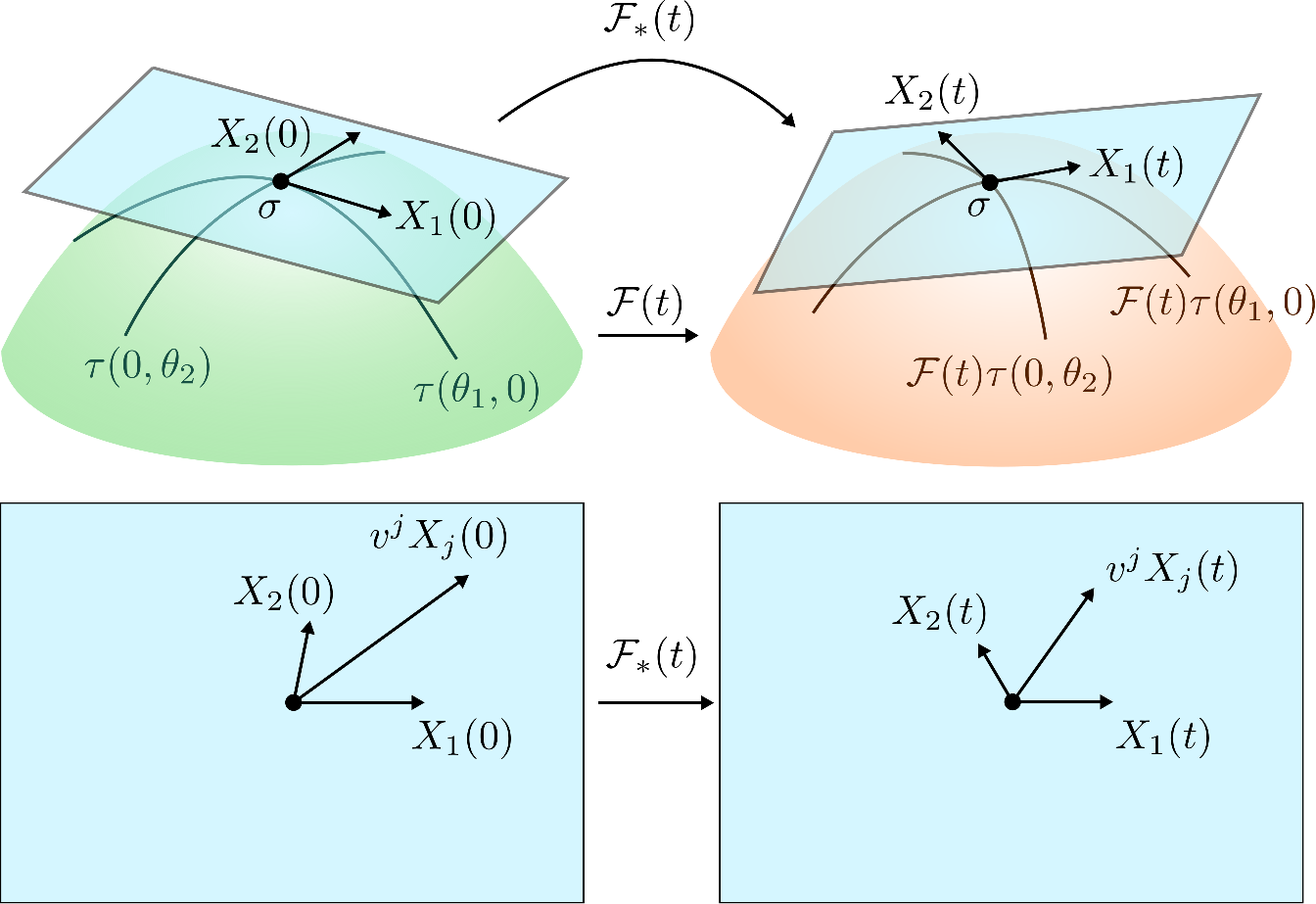}{Top left: the curved surface represents a
  manifold of density operators $\{\tau(\theta):\theta \in \Phi\}$
  that can be prepared for the initial state. The $\tau(\theta_1,0)$
  curve represents the set of density operators for varying $\theta_1$
  and fixed $\theta_2 = 0$; the $\tau(0,\theta_2)$ curve is defined
  similarly. Each score $X_j(0) = \mc E_\sigma^{-1}\partial_j\tau$
  represents the tangent vector of a curve at $\tau(0) = \sigma$. Top
  right: The CPTP map $\mc F(t)$ maps the manifold of density
  operators to a new manifold, while the pushforward $\mc F_*(t)$ maps
  each tangent vector $X_j(0)$ of a curve to the tangent vector
  $X_j(t) = \mc F_*(t) X_j(0)$ of the corresponding curve in the new
  manifold. Bottom left: Given an initial state $\tau(\epsilon v)$,
  $v^jX_j(0)$ represents the tangent vector that points from $\sigma$
  to $\tau(\epsilon v)$. Bottom right: The pushforward $\mc F_*(t)$
  determines the new tangent vectors after time $t$.}

It can be shown that, as time goes on, the expected score with respect
to $\mc F(t)\tau(0) = \mc F(t) \sigma = \sigma$ for any density map
always remains zero, viz.,
\begin{align}
\trace X_j(t) \sigma &= 0\quad\forall t,
\label{Xmean}
\end{align}
and its time evolution can be expressed as
\begin{align}
X_j(t) &= \mc F_*(t) X_j(0),
\label{score_push}
\\
\mc F_*(t) &\equiv \mc E_{\mc F(t)\sigma}^{-1} \mc F(t) \mc E_\sigma
= \mc E_{\sigma}^{-1} \mc F(t) \mc E_\sigma =\exp(\mc L_* t),
\label{retro}
\\
\dot X_j(t) &= \mc L_* X_j(t),
\label{Xrate}
\\
\mc L_* &\equiv \mc E_{\sigma}^{-1} \mc L \mc E_\sigma,
\label{L_push}
\end{align}
where $\mc F_*(t)$ is called a generalized conditional expectation, a
pushforward, or a retrodiction
\cite{hayashi,gce_pra,gce2,parzygnat23}, and $\mc L_*$ is its
generator. 

I stress that $X_j(t)$ is not the Heisenberg picture of
$X_j(0)$. Rather, $X_j(t)$ can be interpreted as the closest
approximation of $X_j(0)$ at time $t$ with respect to a certain
distance measure, as discussed in Refs.~\cite{gce_pra,gce2},
generalizing the least-squares interpretation of the classical
conditional expectation.

An alternative definition of the retrodiction map is as follows.
Let $\mc M^*$ be the Hilbert-Schmidt adjoint of a map $\mc M$, viz.,
\begin{align}
\Avg{A,\mc M B} &= \Avg{\mc M^* A,B}\quad\forall A,B \in \mc O(\mc H).
\label{adjoint}
\end{align}
The Hilbert-Schmidt adjoint $\mc F^*(t)$ of the CPTP map
$\mc F(t)$ is also a generalized conditional expectation, except that
it is predictive \cite{gce2}.  The retrodictive $\mc F_*(t)$ can then
be defined as the adjoint of the predictive $\mc F^*(t)$ with respect
to the weighted inner product, viz.,
\begin{align}
\Avg{\mc F^*(t) A,B}_\sigma &= \Avg{A,\mc F_*(t) B}_\sigma
\quad\forall A,B \in \mc O(\mc H).
\label{weighted_adjoint}
\end{align}
I call $\mc F_*$ the weighted adjoint of $\mc F^*$ in this
mathematical context.  Similarly, let $\mc L^*$ be the Hilbert-Schmidt
adjoint of $\mc L$.  Then $\mc L_*$ is the weighted adjoint of
$\mc L^*$, viz.,
\begin{align}
\Avg{\mc L^* A,B}_\sigma &= \Avg{A,\mc L_*B}_\sigma
\quad\forall A,B \in \mc O(\mc H).
\label{adjoint2}
\end{align}
An illuminating example is the classical case shown in
Appendix~\ref{app_classical}, when all the operators commute. The
quantum case is more complicated because many quantities depend on the
choice of the density map. To clarify, Tables~\ref{table} list the
main quantities in this work and whether each depends the choice of
$\mc E_\sigma$.

\begin{table*}[htbp!]
\begin{tabularx}{\textwidth}{|X|X|X|X|X|X|X|}
\hline
Symbol & $\mc F(t)$ & $\mc L$ & $\sigma$ & $\rho(0) = \tau(\theta)$ & $\theta = \epsilon v$
& $\avg{\cdot,\cdot}_\sigma$  \\
\hline
Meaning & Markov semigroup & generator of $\mc F(t)$ & steady state of $\mc F(t)$ 
& initial state & parameters of initial state &  weighted inner product \\
\hline
Depends on the choice of $\mc E_\sigma$?
& No & No & No & No & No & Yes  \\
\hline
\end{tabularx}

\begin{tabularx}{\textwidth}{|X|X|X|X|X|X|X|}
\hline
Symbol & $J(t)$ & $X_j(t)$ & $\mc F_*(t)$ & $\mc L_*$ & $\mc F^*(t)$ & $\mc L^*$ \\
\hline
Meaning & Fisher information & score & retrodiction & generator of $\mc F_*(t)$ 
& prediction & generator of $\mc F^*(t)$\\
\hline
Depends on the choice of $\mc E_\sigma$? & Yes & Yes &  Yes & Yes & No & No \\
\hline
\end{tabularx}

\caption{\label{table}The main quantities used in this work, their
  meanings, and whether each depends on the choice of the density map
  $\mc E_\sigma$.}
\end{table*}

A vital property of information metrics is monotonicity, which means
that $v^j J_{jk} v^k$ for any $v$ cannot increase after any CPTP map
is applied to $\tau(\theta)$. For a finite-dimensional $\mc O(\mc H)$,
Petz has completely characterized the class of density maps that give
monotone information metrics \cite{petz96,petz}, and I assume such a
map in the following. To be explicit, the information metric is
monotonic if and only if \cite{petz96,petz}
\begin{align}
\mc E_\sigma
&= \mc R_\sigma \phi(\Delta_\sigma),
\label{petz}
\\
\mc R_\sigma A &\equiv A \sigma,
\label{R_map}
\\
\Delta_\sigma A &\equiv \sigma A \sigma^{-1},
\label{Delta}
\end{align}
where $\phi:(0,\infty) \to \mb R$ is an operator monotone function
\cite[Sec.~11.6]{petz}, $\mc R_\sigma$ is called the right product
map, and $\Delta_\sigma$ is called the modular map. In addition,
$\mc E_\sigma$ is symmetric if and only if $\phi$ satisfies
$\phi(u) = u \phi(1/u)$ \cite[Theorem~7]{petz96}. $\phi(1) = 1$ may
also be assumed with little loss of generality to normalize
$\mc E_\sigma$, such that $\mc E_\sigma A = \sigma A$ if $\sigma$ and
$A$ commute.  Define the rate at which the system converges to the
steady state as
\begin{align}
R(t) &\equiv -\dv{}{t} D(\rho(t)\Vert\sigma)
= -\frac{1}{2}\epsilon^2 v^j\dot J_{jk}(t)  v^k + o(\epsilon^2).
\label{R}
\end{align}
Monotonicity of $J$ implies
\begin{align}
v^j \dot J_{jk}(t) v^k \le  0\quad\forall t, v,
\label{J_mono}
\end{align}
and the convergence rate is always nonnegative under the approximation
given by Eq.~(\ref{R}). In other words, monotonicity is a
generalization of the second law of thermodynamics, \new{and the
  Markov semigroup model of the dynamics ensures
  that the second law holds over any time interval.}

The most popular divergence may be the Umegaki relative entropy
\begin{align}
D(\rho\Vert\sigma) &= \trace \rho\bk{\ln \rho-\ln \sigma}.
\label{umegaki}
\end{align}
$R(t)$ is then the commonly assumed entropy production rate in quantum
thermodynamics \cite{alicki_lendi}.  The Umegaki relative entropy can
be approximated by Eq.~(\ref{D}) in terms of the so-called
Kubo-Mori-Bogoliubov (KMB) version of the Fisher information
\cite[Eq.~(6.34)]{hayashi}.  This KMB version is given by
Eq.~(\ref{J}) in terms of the density map \cite[Eq.~(6.9)]{hayashi}
\begin{align}
\mc E_{\sigma,\KMB} A \equiv
 \int_0^1 \sigma^\lambda A \sigma^{1-\lambda} d\lambda,
\label{bogo}
\end{align}
where the second subscript of $\mc E_{\sigma,\mr{version}}$ specifies
the version of the density map \footnote{The weighted inner product in
  terms of Eq.~(\ref{bogo}) is also called the canonical correlation
  \cite{kubo}. Although this inner product is often attributed to
  Kubo, Mori, and Bogoliubov, Ref.~\cite{kubo54} by Kubo and Tomita in
  1954 appears to be the earliest reference. Mori credits
  Ref.~\cite{kubo54} for the formula in his Ref.~\cite{mori56}, while
  some references, such as Ref.~\cite{roepstorff76}, credit a 1962
  paper by Bogoliubov \cite{bogo62}.}.  Another popular divergence
measure is the squared Bures distance \cite{hayashi}
\begin{align}
D(\rho\Vert\sigma) &= 4\bk{1 - \trace\sqrt{\sqrt{\rho}\sigma\sqrt{\rho}}},
\end{align}
which has the useful advantage of remaining finite for pure states and
is important for quantum hypothesis testing. As is well known
\cite{hayashi,petz,uhlmann_crell,braunstein}, it can be approximated
by Eq.~(\ref{D}) \cite[Eq.~(6.33)]{hayashi} using Helstrom's version
of the Fisher information, which is Eq.~(\ref{J}) in terms of the
density map \cite[Eqs.~(6.8) and (6.33)]{hayashi}
\begin{align}
\mc E_{\sigma,\Hel} A \equiv \frac{1}{2}(\sigma A + A \sigma).
\label{Hel}
\end{align}
The Helstrom information plays a fundamental role in quantum
estimation theory \cite{helstrom,hayashi}.  Both the Umegaki relative
entropy and the Bures distance are exactly monotonic, although only
the monotonicity of $J$ is needed in what follows.

For any symmetric and normalized $\mc E_\sigma$ in Petz's class and
any Hermitian operator $A$, Petz has shown that
\cite[Example~10.12]{petz}
\begin{align}
\trace A \mc E_{\sigma,\Hel}^{-1} A
&\le \trace A \mc E_{\sigma}^{-1} A \le 
\trace A \sigma^{-1} A.
\end{align}
These inequalities may be useful for bounding the Fisher information
in case one's preferred version is difficult to compute. For example,
the Helstrom version can be used to lower-bound the KMB version
for a study of entropy production, or the KMB version can be
used to upper-bound the Helstrom version for a study of parameter
estimation.

\begin{example}[continues=exa_gibbs]
  For the Gibbs parametrization given by Eqs.~(\ref{gibbs_sigma}) and
  (\ref{gibbs}), one can use Duhamel's formula \cite[Eq.~(6.32)]{hayashi}
\begin{align}
\partial_j[\exp A(\theta)]
&= \mc E_{\exp A(0),\KMB}\Bk{\partial_j A(\theta)}
\end{align}
to obtain the KMB score
\begin{align}
X_j(0) &= \partial_j \ln \tau = G_j - \trace G_j \sigma.
\end{align}
As $(\theta^j,G_j)$ can be regarded as a pair of conjugate variables
in thermodynamics, $(\epsilon v^j,X_j(0))$ are the deviations of the
conjugate variables from the steady-state values
$(\lambda^j,\trace G_j\sigma)$. 
\end{example}

\begin{example}[continues=exa_kick]
Consider a bosonic field with $s$ modes and $2s$ canonical observables
\begin{align}
Q &\equiv \mqty(q_1 & \dots & q_s& p_1 &\dots & p_s)^\top,
\label{Q}
\end{align}
where $\top$ denotes the matrix transpose such that $Q$ is a column
vector.  The canonical observables satisfy the commutation relation
\begin{align}
[q_j,p_k] &= i\delta_{jk},
\end{align}
where $\delta_{jk}$ denotes the Kronecker delta. The commutation
relation can also be expressed as
\begin{align}
[Q_j,Q_k] &= i \Omega_{jk},
\label{comm}
\end{align}
where $\Omega$ is the $2s\times 2s$ symplectic matrix defined by
\begin{align}
\Omega_{jk} &\equiv 
\begin{cases}
1, & j = 1,\dots,s, k = j+s,
\\
-1, & j = k+s, k = 1,\dots, s,
\\
0, & \textrm{otherwise}.
\end{cases}
\label{Omega}
\end{align}
In matrix form, one can write
\begin{align}
\Omega &= \mqty(0 & I\\ -I & 0),
\end{align}
where $0$ is the $s\times s$ zero matrix and $I$ is the
$s\times s$ identity matrix.  Define also the annihilation operators
\begin{align}
a_j &\equiv \frac{1}{\sqrt{2}} (q_j+ip_j),\quad j = 1,\dots,s,
\end{align}
satisfying
\begin{align}
[a_j,a_k^\dagger] = \delta_{jk}.
\end{align}
For this example, no insight is gained by distinguishing between upper
and lower indices, so I use only lower indices in the following for
clarity, although Einstein summation remains in use.

The Helstrom density map given by Eq.~(\ref{Hel}) is hereafter assumed
in this example. Let the steady state be the thermal state
\begin{align}
\sigma &=
\frac{1}{(\bar n+1)^s}
 \bk{\frac{\bar n}{\bar n+1}}^{a_j^\dagger a_j},
\label{thermal}
\end{align}
where $\bar n$ is the mean particle number in each mode
($a_j^\dagger a_j$ with Einstein summation means
$\sum_j a_j^\dagger a_j$). It can be shown that
\begin{align}
\trace \sigma Q_j &= 0,
& 
\Avg{Q_j,Q_k}_\sigma &= \bk{\bar n+\frac{1}{2}} \delta_{jk}.
\end{align}
Let the initial state $\rho(0)$ be the kicked state given by
Eq.~(\ref{kick}), and let the kick generators in Eq.~(\ref{kick}) be
\begin{align}
G_j &= \Omega_{jk} Q_k.
\label{G_kick}
\end{align}
$\epsilon v$ then quantifies the displacement of the initial state in
phase space. It can be shown using the Gaussian theory in
Appendix~\ref{app_gauss} and in particular Prop.~\ref{prop_sld} that
the scores are given by
\begin{align}
\partial_j\tau &= -i \Omega_{jk} [Q_k,\sigma] = \mc E_{\sigma,\Hel} X_j(0),
\\
X_j(0) &= r Q_j,
\label{score_kick}
\\
r &\equiv \frac{1}{\bar n+1/2}.
\label{r}
\end{align}
The initial Helstrom information becomes
\begin{align}
J_{jk}(0) &= r \delta_{jk}.
\label{J0_kick}
\end{align}
\new{$J(t)$ will be worked out in Sec.~\ref{sec_onsager_geo},
once the dynamics is specified.}
\end{example}

The stage is now set for the core results of this work.

\section{\label{sec_results}Results}
\subsection{\label{sec_onsager}Quantum Onsager theory}

\subsubsection{\label{sec_onsager_geo}Geometric version}
I first prove an analog of Eqs.~(\ref{R_intro})--(\ref{onsager}) that
is natural in the context of quantum information geometry.
\begin{theorem}
\label{thm}
Define a generalized force as
\begin{align}
f^j &\equiv -\epsilon v^j,
\label{f}
\end{align}
and an averaged conjugate variable as
\begin{align}
x_j(t) &\equiv \trace X_j(t) \rho(t),
\label{x}
\end{align}
which I call a geometric variable. The convergence rate is then given
by
\begin{align}
R(t) &= \frac{1}{2}  f^j \dot x_j(t)
+ o(\epsilon^2).
\label{R2}
\end{align}
The geometric current $\dot x(t)$ obeys the equation of motion
\begin{align}
\dot x_j(t) &= K_{jk}(t) f^k + o(\epsilon),
\label{xrate}
\end{align}
where the transport tensor $K$ is the information loss rate given by
\begin{align}
K_{jk}(t) &= -\dot J_{jk}(t)
= -\Avg{X_j(t),\bk{\mc L^*+\mc L_*} X_k(t)}_\sigma.
\label{K}
\end{align}
For any time $t$, $K(t)$ is symmetric, viz.,
\begin{align}
K_{jk}(t) = K_{kj}(t),
\end{align}
and positive-semidefinite, so $K$ can be regarded as a Riemannian
metric.
\end{theorem}
The proof is delegated to Appendix~\ref{app_math}.

By identifying $f = -\epsilon v$ as the force, Theorem~\ref{thm} shows
that, under minimal assumptions, the approximation of the divergence
by the Fisher information as per Eq.~(\ref{D}) can be regarded as the
origin of the standard quadratic approximation of the entropy
production rate in the Onsager theory \cite[Eq.~(2-7)]{onsager53}. The
connection between the Fisher information and the Onsager theory is
hence fundamental, though hitherto unappreciated.  Once the connection
is made, Eq.~(\ref{R2}) for the convergence rate follows; it depends
on the force and a current in the same manner as
Eq.~(\ref{R_intro}). The symmetry of $K$ given by Eq.~(\ref{K}) is
also analogous to the Onsager relation given by Eq.~(\ref{onsager}).

Even though Theorem~\ref{thm} resembles the Onsager theory, it differs
from earlier versions of the theory in the following ways:
\begin{enumerate}[label=(\roman*)]

\item The theorem requires only the assumption of a Markov semigroup,
  the existence of a steady state $\sigma$ as per Eqs.~(\ref{steady}),
  states being close to it in the sense of the Taylor approximation
  given by Eq.~(\ref{taylor}), and a divergence measure given by
  Eq.~(\ref{D}). No other thermodynamic concepts are involved; no
  explicit parametrization of $\tau(\theta)$ is even needed.  On the
  other hand, the literature often assumes Gaussian statistics and
  linear dynamics for the variables
  \cite{reichl,landau_stat,salazar20}, Spohn and Lebowitz made heavy
  assumptions about the whole model based on thermodynamics
  \cite{spohn78}, Alicki assumed an exponential parametrization
  similar to Eq.~(\ref{gibbs}) \cite{alicki76}, and Lendi assumed a
  linear parametrization \cite{lendi86}.

\item Theorem~\ref{thm} is proven for a large class of divergence
  measures and information metrics.
  Refs.~\cite{alicki76,spohn78,lendi86,lendi01,rodriguez20}, for
  example, focus on the Umegaki case.

\item No detailed balance of any kind has been assumed to prove the
  symmetry of the transport tensor $K$---its symmetry is inherited
  from the information metric.

\item Theorem~\ref{thm} holds for time-varying rates at any $t$ and
  the transport tensor there is time-dependent.  Previous studies
  often derive the Onsager relations by considering the rates at
  $t = 0$ only---see, for example, Refs.~\cite[Sec.~7.4]{reichl},
  \cite[Eq.~(V.6)]{spohn78}, \cite[Eq.~(9)]{lendi86}, and
  \cite[Eq.~(306)]{alicki_lendi}---and they require further
  assumptions, such as slow dynamics, to extrapolate the relations to
  longer times.

\item Compared with Eq.~(\ref{R_intro}), the convergence rate equation
  given by Eq.~(\ref{R2}) has an extra factor of $1/2$ on the
  right-hand side.

\end{enumerate}
I emphasize that all these points remain true for the classical case,
except point (ii). Since there is only one classical Fisher metric,
all the classical divergence measures assumed by Eq.~(\ref{D}) end up
having the same leading order.

The physical meaning of the geometric variable $x(t)$ comes from
Eq.~(\ref{R2})---its rate $\dot x(t)$ is precisely the covector that
determines the convergence rate at all times, in the same manner as
Eq.~(\ref{R_intro}) in the original Onsager theory. To measure
$x_j(t)$, one should measure the averaged variable given by
Eq.~(\ref{x}), which is in terms of an observable $X_j(t)$ with an
explicit time dependence. A von Neumann measurement of $X_j(t)$ at
each time is modeled by the projection-valued measure $\Pi(\lambda,t)$
in the spectral resolution
$X_j(t) = \sum_\lambda \lambda \Pi(\lambda,t)$, and $\Pi(\lambda,t)$
here is time-varying and may be difficult to perform.  However,
$x_j(t)$ is only the average of $X_j(t)$, and there is more freedom in
how the average can be estimated, as illustrated by the following
example.
\begin{example}[continues=exa_kick]
  Suppose that $\mc F(t)$ is a Gaussian channel \cite{holevo19}.  With
  each initial score $X_j(0)$ given by Eq.~(\ref{score_kick}), which
  is proportional to a canonical observable,
  Ref.~\cite[Proposition~1]{gce2} has shown that $\mc F_*(t) X_j(0)$
  given by Eq.~(\ref{score_push}) remains a linear function of the
  canonical observables, viz.,
\begin{align}
X_j(t) =  c_j(t) + C_{jk}(t) Q_k
\end{align}
for some $c(t) \in \mb R^{2s}$ and $C(t) \in \mb R^{2s\times 2s}$; see
also Prop.~\ref{prop_gce} in Appendix~\ref{app_gauss}. The expected
value becomes
\begin{align}
x_j(t) = \trace X_j(t)\rho(t) = c_j(t) + C_{jk}(t) \trace Q_k \rho(t).
\label{x_linear}
\end{align}
If the field is optical, then each $X_j(t)$ is a time-varying
quadrature, and the expected value can be estimated, for example, by
heterodyne detection of all the relevant quadratures $\{Q_k\}$ over
many trials, followed by averaging and data processing in accordance
with Eq.~(\ref{x_linear}). 

Since $\mc F^*$, $\mc F_*$, $\mc L^*$, and $\mc L_*$ for a Gaussian
channel can be shown to map canonical observables to canonical
observables, while the scores are also linear functions of canonical
observables, it suffices in the following to consider their effects on
$Q$ only, rather than the full $\mc O(\mc H)$ or $\mc O_s(\mc H)$.

To give a more concrete example of a Gaussian channel, suppose that
the system consists of passively coupled and damped harmonic
oscillators, such that the semigroup generator is given by
\cite{gardiner_zoller}
\begin{align}
\mc L\rho &= -i \omega_{jk} \Bk{a_j^\dagger a_k,\rho}
+ \frac{1}{2}(\bar n+1)\gamma_{jk}  (2 a_k \rho a_j^\dagger
- a_j^\dagger a_k \rho - \rho a_j^\dagger a_k)
+ \frac{1}{2}\bar n\gamma_{jk}
(2 a_j^\dagger \rho a_k - a_k a_j^\dagger \rho - \rho a_k a_j^\dagger),
\label{GKSL}
\end{align}
where $\omega \in \mb C^{s\times s}$ and
$\gamma \in \mb C^{s\times s}$ are complex Hermitian matrices, viz.,
\begin{align}
\omega &= \omega^\dagger,
&
\gamma &= \gamma^\dagger.
\end{align}
Moreover, $\gamma$ must be positive-semidefinite so that $\mc L$ is in
the GKSL form.  It can be shown that the thermal state given by
Eq.~(\ref{thermal}) is a steady state; see Prop.~\ref{prop_thermal} in
Appendix~\ref{app_thermal}. It can also be shown that the $\mc L_*$
map for this example turns out to be the same for any density map in
Petz's class; see Props.~\ref{prop_Delta} and \ref{prop_commute} in
Appendix~\ref{app_thermal}.

With Eq.~(\ref{GKSL}), the equation of motion for the mean field
$\bar a_j(t) \equiv \trace a_j \rho(t)$ becomes
\begin{align}
\dv{\bar a_j}{t}&= \Lambda_{jk} \bar a_k,
&
\Lambda_{jk} &\equiv -i \omega_{jk}  - \frac{1}{2} \gamma_{jk}.
\end{align}
This form of coupled-mode equations is well known in classical optics,
where $\omega$ governs the conservative coupling between the optical
modes and $\gamma$ governs the loss. The equation of motion for
$\bar Q_j(t) \equiv \trace Q_j \rho(t)$ becomes
\begin{align}
\dv{\bar Q_j}{t} &= L_{jk} \bar Q_k
\end{align}
with a certain rate matrix $L \in \mb R^{2s\times 2s}$. If $\bar a$
and $\bar Q$ are regarded as column vectors, then $L$ is related to
$\Lambda$, $\omega$, and $\gamma$ by
\begin{align}
L &= \mqty(\Re\Lambda  & -\Im\Lambda\\ 
\Im\Lambda & \Re\Lambda)
= \mqty(\Im\omega -\Re\gamma/2 & \Re\omega+\Im\gamma/2\\
-\Re\omega-\Im\gamma/2 & \Im\omega -\Re\gamma/2),
\label{L}
\end{align}
where $\Re$ denotes the entry-wise real part and $\Im$ denotes the
entry-wise imaginary part. Defining
\begin{align}
F(t) &\equiv \exp(L t),
\end{align}
assuming the thermal steady state given by Eq.~(\ref{thermal}), and
using Appendix~\ref{app_gauss} and in particular Prop.~\ref{prop_gce},
it can be shown that
\begin{align}
\mc F^*(t) Q_j &= F_{jk}(t) Q_k,
&
\mc F_*(t) Q_j &= F_{kj}(t) Q_k,
\end{align}
or, in matrix form,
\begin{align}
\mc F^*(t) Q &= F(t) Q,
&
\mc F_*(t) Q &= F^\top(t) Q,
&
\mc L^* Q &= L Q,
&
\mc L_* Q &= L^\top Q,
\end{align}
where $\mc M Q$ for a map $\mc M$ is understood as the column vector
$\mqty(\mc M Q_1 & \mc M Q_2 & \dots)^\top$. Starting with
Eq.~(\ref{score_kick}), the scores become
\begin{align}
X(t) &= \mc F_*(t) X(0) = r F^\top(t) Q,
\label{Xt}
\end{align}
while the Helstrom information matrix and the information loss rate
are respectively given by
\begin{align}
J(t) &= rF^\top(t) F(t),
\\
K(t) &= -r F^\top(t)\bk{ L^\top  +  L }F(t).
\label{K_kick}
\end{align}
\end{example}
\subsubsection{Physical version}
Theorem~\ref{thm} may seem abstract, but it can be used to prove a
more physical generalization of the Onsager equations given by
Eqs.~(\ref{R_intro}) and (\ref{rate_intro}) in the next theorem;
points (i) and (ii) remain true for the following theorem, while
points (iii)--(v) no longer apply.
\begin{theorem}
\label{thm2}
Define
\begin{align}
y_j(t) &\equiv \trace X_j(0) \rho(t),
\label{y}
\end{align}
which I call an experimental variable. Eq.~(\ref{y}) differs from the
geometric version given by Eq.~(\ref{x}) in assuming that each
observable $X_j(0)$ is fixed in time. Its current at $t = 0$ obeys
\begin{align}
\dot y_j(0) &= O_{jk}f^k  +o(\epsilon),
\label{yrate}
\end{align}
where the Onsager transport tensor $O$ is given by
\begin{align}
O_{jk} &= -\Avg{X_j(0),\mc L_* X_k(0)}_\sigma.
\label{O}
\end{align}
$O$ is related to the information loss rate $K$ defined in
Eq.~(\ref{K}) by
\begin{align}
K_{jk}(0) &= O_{jk} + O_{kj},
\label{KO}
\end{align}
and $O$ is positive-semidefinite.  The convergence rate becomes
\begin{align}
R(0) &= f^j \dot y_j(0)  + o(\epsilon^2).
\label{R3}
\end{align}
\end{theorem}
The proof is delegated to Appendix~\ref{app_math}.

Like Theorem~\ref{thm}, Theorem~\ref{thm2} makes minimal assumptions
about the system and is valid for arbitrary parametrizations of the
initial state and a large class of divergence measures and information
metrics. The important feature of Theorem~\ref{thm2} is its more
physical definition of the currents, so that it is closer to
conventional experimental settings.  \new{Theorem~\ref{thm2} in itself
  does not require any detailed-balance condition to hold, but it also
  does not prove any symmetry of the transport tensor $O$ given by
  Eq.~(\ref{O}); a generalization of the Onsager relations for the $O$
  tensor must wait until Sec.~\ref{sec_casimir}, where quantum
  detailed-balance conditions are introduced.} Moreover,
  Theorem~\ref{thm2} holds at only one time, meaning that it is
  relevant only to systems with near-static dynamics.

The mysterious factor of $1/2$ in Eq.~(\ref{R2}) in Theorem~\ref{thm}
is absent from Eq.~(\ref{R3}) in Theorem~\ref{thm2}, because the
experimental current $\dot y$ does not include the rate due to
$\dot X_j$ in the geometric version, which comes from the explicit
time dependence of $X_j(t)$.  Omitting the $o(\epsilon^n)$ terms for
brevity, Eqs.~(\ref{R2}) and (\ref{R3}) can be reconciled at $t = 0$
by noting that
\begin{align}
\frac{1}{2} f^j \dot x_j(0) = f^j \dot y_j(0),
\end{align}
even though $\dot x_j(0)/2 = K_{jk}(0)f^k/2$ and
$\dot y_j(0) = O_{jk}f^k$ may be different. An intuitive viewpoint, as
expounded in the classical literature \cite{bertini15}, is to think of
\begin{align}
\frac{1}{2} \dot x_j(0) &= \frac{1}{2} \bk{O_{jk} + O_{kj}} f^k
\end{align}
as the dissipative component of the current $\dot y_j(0)$ that
contributes to the convergence rate, and their difference
\begin{align}
\dot y_j(0) - \frac{1}{2} \dot x_j(0) = \frac{1}{2}\bk{O_{jk}-O_{kj}} f^k
\end{align}
as the conservative component of the current that does not, since the
latter is orthogonal to the force, viz.,
\begin{align}
f^j\Bk{\dot y_j(0) - \frac{1}{2}\dot x_j(0)} = 0.
\end{align}

\begin{example}[continues=exa_gibbs]
  Assume the Gibbs state given by Eq.~(\ref{gibbs}) and let
\begin{align}
G &= \mqty(E_1 & \dots & E_s & N_1 & \dots & N_s)^\top
\end{align}
be a vector of energy operators $\{E_j\}$ and particle-number
operators $\{N_j\}$ of $s$ subsystems. Write the generalized forces as
\begin{align}
f = -\epsilon v = 
\mqty(\delta\beta^1 & \dots & \delta\beta^s
& \delta\mu^1 & \dots & \delta\mu^s)^\top,
\end{align}
which are perturbations to the inverse temperatures
$\{\delta\beta^j\}$ and the (negative and normalized) chemical
potentials $\{\delta\mu^j\}$ of the subsystems. Treating both $G$ and
$f$ as column vectors, the Onsager matrix can be partitioned as
\begin{align}
O &= \mqty(O^{(E,\beta)} & O^{(E,\mu)}\\ O^{(N,\beta)} & O^{(N,\mu)}),
\end{align}
where each $O^{(G,f)}$ submatrix models the response of the currents
$\{\trace G_j\dot\rho(0)\}$ to the forces $\{f^j\}$. The Onsager
symmetry given by Eq.~(\ref{onsager}) would imply
\begin{align}
O_{jk}^{(E,\beta)} &= O_{kj}^{(E,\beta)},
&
O_{jk}^{(N,\mu)} &= O_{kj}^{(N,\mu)},
&
O_{jk}^{(E,\mu)} &= O_{kj}^{(N,\beta)}.
\end{align}
The last relation is a classic example of the Onsager relation,
showing that the transport coefficients $O^{(E,\mu)}$ for the energy
currents due to $\delta\mu$ are related to the coefficients
$O^{(N,\beta)}$ for the particle currents due to $\delta\beta$.
\end{example}

\begin{example}[continues=exa_kick]
  For the kicked initial state given by Eq.~(\ref{kick}), the kick
  generators given by Eq.~(\ref{G_kick}), and the dynamics governed by
  Eq.~(\ref{GKSL}), it can be shown by following the previous
  discussion in this example that the Onsager matrix is given by
\begin{align}
O &= -r L,
\label{O_kick}
\end{align}
where $r$ is given by Eq.~(\ref{r}) and $L$ is the rate
matrix given by Eq.~(\ref{L}). The information loss rate $K$ given by
Eq.~(\ref{K_kick}) at $t = 0$ becomes
\begin{align}
K(0) &= -r (L^\top + L),
\label{K0}
\end{align}
which is indeed equal to $O + O^\top$.  Since $\omega$ and $\gamma$
are Hermitian, $\Re\omega$ and $\Re\gamma$ are symmetric while
$\Im\omega$ and $\Im\gamma$ are antisymmetric.  The information loss
rate at $t = 0$ becomes
\begin{align}
K(0) &= O + O^\top =
-r(L + L^\top) = 
r \mqty(\Re \gamma & -\Im\gamma \\ \Im\gamma & \Re\gamma),
\end{align}
which depends only on the $\gamma$ matrix in the dissipative part of
the semigroup generator $\mc L$.
\end{example}

\subsection{\label{sec_casimir}Quantum Onsager-Casimir relations}
Theorem~\ref{thm2} generalizes Eqs.~(\ref{R_intro}) and
(\ref{rate_intro}) in the Onsager theory; the remaining task is to
generalize the Onsager relation given by Eq.~(\ref{onsager}).  As
Eq.~(\ref{onsager}) is often regarded as a consequence of detailed
balance and detailed balance is in turn most commonly associated with
time-reversal symmetry \cite{gardiner}, I begin with a review of the
standard treatment of time reversal in quantum mechanics; see, for
example, \cite[Sec.~4.4]{sakurai}.

The treatment is centered on the definition of an antiunitary operator
$\vartheta : \mc H \to \mc H$ and a motion-reversal map
$\Theta:\mc O(\mc H)\to\mc O(\mc H)$ given by
\begin{align}
\Theta A &\equiv \vartheta A \vartheta^{-1}.
\end{align}
$\Theta$ can be shown to be antiunitary with respect to the
Hilbert-Schmidt inner product, viz.,
$\avg{\Theta A,\Theta B} = \avg{B,A}$ for any $A,B \in \mc O(\mc
H)$. A Hilbert-Schmidt adjoint $\Theta^*$ of $\Theta$ can be defined
by $\Avg{A,\Theta B} = \Avg{B,\Theta^* A}$ for any $A,B$, and one has
$\Theta^* = \Theta^{-1}$ because of the antiunitarity. It can also be
shown that $\Theta$ maps Hermitian operators to Hermitian operators. A
standard example is a $\Theta$ that gives $\Theta q_j = q_j$ for a
position operator $q_j$ and $\Theta p_j = -p_j$ for a momentum
operator $p_j$. 

A Hamiltonian $H$ is said to satisfy time-reversal symmetry if
$\Theta H = H$. The Heisenberg picture illustrates the concept: if the
evolution from an operator $A(0)$ to another operator $A(t)$ after
time $t$ is described by the equation of motion
\begin{align}
A(t) &= U^\dagger(t) A(0) U(t),
&
U(t) &\equiv \exp(-i H t),
\end{align}
then reversing the final $A(t)$ by $\Theta$ and setting $\Theta A(t)$
as the initial condition of the dynamics would lead to $\Theta A(0)$
after time $t$, because
\begin{align}
\Theta A(0) = U^\dagger(t) [\Theta A(t)] U(t).
\end{align}
Even if a system does not satisfy time-reversal symmetry, it is often
possible to prepare another system with a Hamiltonian given by
$\Theta H$, by reversing an external classical magnetic field for
example. Defining the reverse unitary as
\begin{align}
V(t) \equiv \exp(-i\Theta H t),
\label{V}
\end{align}
one obtains the reversed equation of motion
\begin{align}
\Theta A(0) &= V^\dagger(t) [\Theta A(t)] V(t)
\label{heisenberg2}
\end{align}
under the Heisenberg picture of the reverse unitary.

In Refs.~\cite{onsager31b,casimir45}, Onsager and Casimir considered
systems under external fields that change their signs under motion
reversal, such as a magnetic field or a Coriolis force. They proposed
that, upon reversing the external fields, the new Onsager tensor
$\tilde O$ should be related to the original $O$.  To generalize the
Onsager-Casimir relations to the quantum case, I need to generalize
the concept of time reversal and the reverse unitary given by
Eq.~(\ref{V}) for open systems.

Assume a second open system with dynamics modeled by a quantum
dynamical semigroup $\{\mc G(t) = \exp(\mc K t): t\ge 0\}$. Inspired
by Ref.~\cite{reversal}, I say that $\mc G$ is the reverse of the
$\mc F$ map of the first system if they satisfy
\begin{align}
\mc G^* &= \Theta \mc F_* \Theta^*,
\label{reverse_map}
\end{align}
which is equivalent to 
\begin{align}
\mc K^* &= \Theta \mc L_* \Theta^*
\label{reverse_gen}
\end{align}
for the semigroup generators. To see why this definition makes sense,
observe that it generalizes the reverse unitary given by
Eq.~(\ref{V}): Given $\mc F A = U A U^\dagger$, one obtains
\begin{align}
\mc F^* A &= U^\dagger A U,
&
\mc F_* A &= U A U^\dagger,
&
\mc G^* A &= V^\dagger A V.
\end{align}
One can also arrive at Eq.~(\ref{reverse_map}) by assuming the reverse
unitary for the total system including the reservoir on a larger
Hilbert space---see Appendix~\ref{app_reverse}. In general, the
right-hand side of Eq.~(\ref{reverse_map}) may not be completely
positive, but it is guaranteed to be so if the density map is given by
the so-called Connes version \footnote{Eq.~(\ref{connes}) is
  attributed to Connes because the resulting weighted inner product
  coincides with the selfpolar form introduced by Connes in
  Ref.~\cite{connes74}; see also Ref.~\cite[Eq.~(8.17)]{ohya}.}
\begin{align}
\mc E_{\sigma,\Con} A &\equiv \sigma^{1/2} A \sigma^{1/2},
\label{connes}
\end{align}
since $\mc F_* = \mc E_{\sigma,\Con}^{-1} \mc F \mc E_{\sigma,\Con}$
and thus $\Theta \mc F_*\Theta^*$ can be expressed in terms of Kraus
operators \cite{reversal}. \new{}
A form of time reversal proposed by Crooks
\cite{crooks08} and generalized by Manzano \textit{et al.}\
\cite{manzano15,manzano18} can be shown to be a special case of
Eq.~(\ref{reverse_map}) in terms of Eq.~(\ref{connes}); see
Appendix~\ref{app_crooks} for details. Prop.~\ref{prop_Delta} in
Appendix~\ref{app_thermal} may also be used to check whether $\mc F_*$
turns out to be the same regardless of the chosen density map and
there is no ambiguity about the reversal relation.  The following
results work for any density map and do not require $\mc G$ to be
completely positive, but $\mc G$ must, of course, be a good model of
physical dynamics for the results to be useful.

Let $\tilde\sigma$ be a steady state of the second system satisfying
$\mc G\tilde\sigma = \tilde\sigma$ and $\mc K\tilde\sigma = 0$.  Given
the reversal relation, it is straightforward to show that
\begin{align}
\tilde\sigma &= \Theta \sigma
\label{steady2}
\end{align}
is a steady state of the second system. Assuming
Eqs.~(\ref{reverse_map}) and (\ref{steady2}), the first system is also
the reverse of the second, since $\sigma = \Theta^* \tilde\sigma$ and
\begin{align}
\mc F^* &= \Theta^* \mc G_*\Theta,
&
\mc G_* &\equiv \mc E_{\tilde\sigma}^{-1}\mc G \mc E_{\tilde\sigma},
\end{align}
as shown by Prop.~\ref{prop_reverse_map} in Appendix~\ref{app_math}.

Let $\tilde\tau:\Phi \to \mc P(\mc H)$ be a parametrization of the
initial state of the second system with
$\tilde\tau(0) = \tilde\sigma$. The scores are defined by
\begin{align}
\partial_j \tilde\tau &= \mc E_{\tilde\sigma} \tilde X_j(0).
\end{align}
Following Theorem~\ref{thm2}, the Onsager tensor for the second system
can be defined as
\begin{align}
\tilde O_{jk} &\equiv -\Avg{\tilde X_j(0),\mc K_* \tilde X_k(0)}_{\tilde\sigma},
&
\mc K_* &\equiv \mc E_{\tilde\sigma}^{-1} \mc K \mc E_{\tilde\sigma}.
\label{O2}
\end{align}
A quantum Onsager-Casimir relation for the two systems can now be
given.

\begin{theorem}
\label{thm3}
Suppose that two systems are reverses of each other
in the sense of Eqs.~(\ref{reverse_gen}) and (\ref{steady2}).
Suppose, furthermore, that the parametrization
of the second system is related to that of the first by
\begin{align}
\tilde\tau(\theta) &= \Theta \tau(\theta)
\label{tau2}
\end{align}
in a neighborhood of $\theta = 0$, such that the scores are related by
\begin{align}
\tilde X_j(0) &= \Theta X_j(0).
\label{X2}
\end{align}
Then their Onsager tensors given by Eqs.~(\ref{O}) and (\ref{O2}) obey
\begin{align}
O_{jk} &= \tilde O_{kj}.
\label{casimir}
\end{align}
\end{theorem}
The proof is delegated to Appendix~\ref{app_math}.

Since it may be difficult to prepare a second system that is the
reverse of the first in everything, including the initial-state
preparation, I offer a variation of Theorem~\ref{thm3} for two systems
with identical steady states and initial-state preparations as
follows.
\begin{theorem}
\label{thm4}
  Suppose that two systems are reverses of each other in the sense of
  Eqs.~(\ref{reverse_gen}) and (\ref{steady2}). Suppose, furthermore,
  that the steady states and the initial-state preparations are
  identical, viz.,
\begin{align}
\tilde\sigma &= \Theta \sigma = \sigma,
&
\tilde\tau(\theta) &= \tau(\theta),
&
\tilde X_j(0) &= X_j(0).
\label{identical}
\end{align}
If the scores transform as
\begin{align}
\Theta X_j(0) &= T_j{}^k X_k(0),
\quad
T \in \mb R^{p\times p},
\label{T}
\end{align}
then the two Onsager tensors are related by
\begin{align}
O_{jk} &= T_k{}^l \tilde O_{lm} T_j{}^m.
\end{align}
\end{theorem}
The proof is delegated to Appendix~\ref{app_math}.

The assumption given by Eqs.~(\ref{identical}) in Theorem~\ref{thm4}
means that the only difference between the two systems is the dynamics
governed by the $\mc F$ and $\mc G$ maps, so that it can model a
device before and after an external field is reversed with all the
other settings intact. Theorem~\ref{thm4} is thus closer in spirit to
the original Onsager-Casimir relations, though less general than
Theorem~\ref{thm3}.  Eq.~(\ref{T}) is a generalization of the common
assumption that observables are either even ($\Theta A = A$) or odd
($\Theta A = -A$) under motion reversal.

A quantum detailed-balance condition can now be defined by positing
that a system is its own reverse, with
\begin{align}
\mc G(t) &= \mc F(t), & \mc K &= \mc L,
& \tilde\sigma &= \sigma,
\label{own_reverse}
\end{align}
such that 
\begin{align}
\mc L^* &= \Theta \mc L_* \Theta^*,
&
\sigma &= \Theta\sigma.
\label{DB2}
\end{align}
Eqs.~(\ref{DB2}) coincide with the standard definition of detailed
balance in the classical case---see Eqs.~(\ref{CDB}) in
Appendix~\ref{app_classical}. By generalizing an argument of
Carmichael and Walls \cite{carmichael76}, Appendix~\ref{app_reverse}
shows that time-reversal symmetry of the total system on a larger
Hilbert space, together with some additional assumptions, can lead to
Eqs.~(\ref{DB2}). Eqs.~(\ref{DB2}) have the same form as the
detailed-balance condition proposed by Agarwal \cite{agarwal73} if one
assumes the density map $\mc E_\sigma = \mc R_\sigma$ given by
Eq.~(\ref{R_map}).  Eqs.~(\ref{DB2}) also give the so-called
SQDB-$\theta$ condition proposed by Fagnola and Umanit\`a
\cite{fagnola10} if the Connes density map given by Eq.~(\ref{connes})
is assumed; Roberts \emph{et al.}\ have worked out many examples that
satisfy the SQDB-$\theta$ condition \cite{roberts20,roberts21}.  It
should be noted that a plethora of other quantum detailed-balance
conditions have been proposed in the literature
\cite{alicki76,kossa77,majewski98,fagnola07,fagnola10,temme10,carlen17}
but it is outside the scope of this work to study those---the version
here is chosen because of its connections with time-reversal symmetry
and the Onsager-Casimir relations.

Under the detailed-balance condition given by Eqs.~(\ref{DB2}),
Theorem~\ref{thm3} becomes a statement about the Onsager tensors for
the same system with two different initial-state preparations, while
Theorem~\ref{thm4} implies a symmetry of the Onsager tensor.
\begin{corollary}
\label{cor_onsager}
Assume that a system satisfies the detailed-balance condition given by
Eqs.~(\ref{DB2}) and the scores transform as Eq.~(\ref{T}) in terms of
a matrix $T$ under motion reversal. Then the Onsager tensor satisfies
\begin{align}
O_{jk} &= T_k{}^l O_{lm} T_j{}^m.
\label{symmetry3}
\end{align}
If $T$ is the Kronecker delta, meaning that
\begin{align}
\Theta X_j &= X_j,
\end{align}
then the original Onsager relation given by Eq.~(\ref{onsager}) is
obtained.
\end{corollary}
\begin{proof}
  Under the detailed-balance condition, $\tilde O = O$ if one assumes
  the same initial-state preparation. Theorem~\ref{thm4} then implies
  Eq.~(\ref{symmetry3}).
\end{proof}

\begin{example}[continues=exa_kick]
Assume a $\Theta$ that satisfies
\begin{align}
\Theta &= \Theta^*,
&
\Theta q_j &= q_j,
&
\Theta p_j &= -p_j.
\end{align}
The map can also be expressed in matrix form
\begin{align}
\Theta Q &= T Q,
&
T &\equiv \mqty(I & 0\\ 0 & -I).
\end{align}
Assume a second system that consists of $s$ bosonic modes with the
same GKSL form given by Eq.~(\ref{GKSL}) for $\mc K$, except that the
matrices $\omega$ and $\gamma$ are denoted by $\tilde\omega$ and $\tilde\gamma$
instead.  The rate matrix for the second system can be defined by
\begin{align}
\mc K^* Q &= \tilde L Q, 
&
\tilde L &\equiv \mqty(\Im\tilde\omega - \Re\tilde\gamma/2 & \Re\tilde\omega+\Im\tilde\gamma/2\\
-\Re\tilde\omega-\Im\tilde\gamma/2 & \Im\tilde\omega-\Re\tilde\gamma/2),
\end{align}
in the same form as the $L$ matrix given by Eq.~(\ref{L}) for the first.
The reversal condition given by Eq.~(\ref{reverse_gen}) would imply
\begin{align}
\tilde L &= T L^\top T = \mqty(-\Im\omega -\Re\gamma/2 & \Re\omega - \Im\gamma/2\\
-\Re\omega + \Im\gamma/2 & -\Im\omega - \Re\gamma/2),
\end{align}
which is satisfied if
\begin{align}
\Re\omega &= \Re\tilde\omega, & \Im\omega &= -\Im\tilde\omega,
&
\Re\gamma &= \Re\tilde\gamma, & \Im\gamma &= -\Im\tilde\gamma,
\end{align}
meaning that $\omega$ and $\tilde\omega$ should be the entry-wise complex
conjugate of each other; likewise for $\gamma$ and $\tilde\gamma$. This
relation is reminiscent of the effect of reversing an external
magnetic field on the permittivity matrix of a magneto-optic medium
\cite{asadchy20}, although a more careful study is needed to ascertain
the connection.

Suppose that the steady state $\tilde\sigma$ of the second system is
the thermal state given by Eq.~(\ref{thermal}). One has
\begin{align}
\tilde\sigma &= \sigma = \Theta\sigma.
\end{align}
Suppose also that the scores of the second system are the reverse of
Eq.~(\ref{score_kick}), viz.,
\begin{align}
\tilde X(0) &= \Theta X(0) = r T Q.
\label{X_kick}
\end{align}
The second Onsager matrix becomes
\begin{align}
\tilde O &= -r T \tilde L T
= -r L^\top = O^\top,
\end{align}
in accordance with Theorem~\ref{thm3}. If the initial-state
preparations are identical, then the scores become
\begin{align}
\tilde X(0) &= X(0) = r Q
\end{align}
instead of Eq.~(\ref{X_kick}), and the second Onsager matrix becomes
\begin{align}
\tilde O &= -r \tilde L = -r  T L^\top T = T O^\top T,
\end{align}
in accordance with Theorem~\ref{thm4}.

Now assume the detailed-balance condition given by
Eqs.~(\ref{DB2}) for the first system, implying
\begin{align}
L &= T L^\top T,
\label{L_sym2}
\end{align}
or equivalently
\begin{align}
\Im\omega &= 0, &  \Im \gamma &= 0,
\end{align}
meaning that all entries of $\omega$ and $\gamma$ are real.  The
Onsager matrix then satisfies
\begin{align}
O &= T O^\top T,
\end{align}
in accordance with Corollary~\ref{cor_onsager}.

\end{example}

\section{\label{sec_discuss}Discussion}
Using the Fisher-information approach put forth, one can treat any
parameter of the initial state as a generalized force, and the
conjugate variables that determine the entropy production rate are
automatically given by the scores, which generalize the concept of
macroscopic state variables in thermodynamics. This approach allows
one to study entropy production with arbitrary perturbations that may
not be well described by traditional thermodynamic concepts. For
example, the concept of Gibbs ensemble, temperature, and chemical
potential in Example~\ref{exa_gibbs} may be ill-defined for quantum
systems under fast control; one may instead adopt a more sophisticated
model of the initial state as a function of the experimental control
parameters.

The connection between the entropy production rate and the Fisher
information loss rate is another discovery that may broaden the
utility of the Onsager theory and enable more cross-pollination
between thermodynamics and parameter estimation theory. The Helstrom
information, for example, plays a fundamental role for important
applications in quantum metrology, such as gravitational-wave
detection \cite{braginsky,twc,danilishin19} and optical imaging
\cite{review_cp}; its loss in open quantum systems is an important
problem under active research
\cite{escher,demkowicz,tsang_open,zhou18}. Can the Onsager theory
offer new insights about how information flows in a quantum sensor and
how information loss can be mitigated, or can quantum metrology bring
new ideas or new testbeds to the area of quantum thermodynamics?

By adopting the concept of density maps pioneered by Petz, the results
put forth are valid for a large class of divergence measures and
information metrics, unlike many prior works that focus on specific
versions \cite{alicki76,spohn78,lendi86,lendi01,rodriguez20}. Even if
one is interested in only one version of divergence, other versions
may still be useful for bounding it if they are easier to compute.

Theorem~\ref{thm} reveals a geometric form of the Onsager theory that
is remarkable in its generality but admittedly obscure in its physical
relevance. Theorem~\ref{thm2} offers a more physical version derived
from Theorem~\ref{thm}, while Theorems~\ref{thm3} and \ref{thm4}
establish quantum Onsager-Casimir relations based on a general concept
of time reversal and detailed balance for open quantum systems. The
many versions of time reversal and detailed balance depending on the
density map, though bewildering, may also be regarded as intriguing
features of the quantum theory---they suggest that one must handle the
concept of time reversal and detailed balance with care in the quantum
regime.

Examples~\ref{exa_gibbs} and \ref{exa_kick} have offered two basic
scenarios to illustrate the core concepts, but there should be a lot
more problems that can benefit from the proposed formalism, given its
generality. It is noteworthy that some quantum Onsager-Casimir
relations for spintronics have been proposed by Jacquod \emph{et al.}\
\cite{jacquod12}, but their approach is very different from Onsager's,
does not involve entropy at all, and seems specific to electronic
systems. Optics is another area where the Onsager-Casimir relations
are of recent interest, as they have been proposed as the origin of
electromagnetic reciprocity \cite{asadchy20}. Given the confusion
regarding reciprocity in lossy optical devices
\cite{fan12,time_bandwidth_ol}, a quantum treatment to ensure
conformance with both quantum mechanics and thermodynamics may be
worthwhile. It is an interesting open question how the formalism here
can be applied to specific areas.

\section{\label{sec_conclusion}Conclusion}
This work has established Fisher information geometry as the
appropriate foundation of the Onsager theory, thereby revealing a
remarkable and fundamental connection between statistical mechanics
and parameter estimation theory.

Despite the significant generalizations made in this work, the Onsager
theory remains limited in its ability to deal with
far-from-equilibrium conditions, time-dependent perturbations, and
fast dynamics. Such situations can, of course, be modeled by other
methods in statistical mechanics \cite{weiss} and are also of
significant interest to quantum metrology
\cite{braginsky,twc,tsang_open,guta11,guta15,guta17,gammelmark14}. The
intersection of the two areas is a fertile ground awaiting further
explorations.

\section*{Acknowledgments}
This research was supported by the National Research Foundation,
Singapore, under its Quantum Engineering Programme (QEP-P7).
Correspondence with Mark Mitchison and the hospitality of Alex
Lvovsky, the Keble College, and the Department of Physics at the
University of Oxford, where part of this work was performed, are
gratefully acknowledged.




\appendix

\section{\label{app_classical}Classical case}

Commuting Hermitian operators can be diagonalized with respect to the
same orthonormal basis of $\mc H$; let such a basis be
$\{\ket{z}\}$. Let $\{Z(t):t \ge 0\}$ be a classical
Markov process with conditional distribution $P_{Z(t)|Z(0)}(z|z')$ and
rate matrix
\begin{align}
r(z|z') \equiv \dot P_{Z(t)|Z(0)}(z|z').
\end{align}
$Z(t)$ can be regarded as the trajectory of all the microscopic
variables in thermodynamics.  Let
\begin{align}
\rho &= \sum_z P(z)\ket{z}\bra{z}
\end{align}
be a generic density operator in terms of a probability distribution
$P$ (Einstein summation is abandoned in this appendix for clarity).
The semigroup can be expressed as
\begin{align}
\mc F(t) \rho &= \sum_{z,z'} P_{Z(t)|Z(0)}(z|z')
P(z')\ket{z}\bra{z},
\\
\mc L \rho &= \sum_{z,z'} r(z|z')P(z')\ket{z}\bra{z}.
\end{align}
Let a steady state be 
\begin{align}
\sigma = \sum_z S(z)\ket{z}\bra{z}
\end{align}
with a distribution $S$ that satisfies
\begin{align}
\sum_{z'} P_{Z(t)|Z(0)}(z|z')S(z') &= S(z).
\end{align}
In terms of generic observables
\begin{align}
A &= \sum_z a(z) \ket{z}\bra{z},
&
B &= \sum_z b(z) \ket{z}\bra{z},
\end{align}
where $a$ and $b$ are real classical random variables, one can write
\begin{align}
\mc E_\sigma A &= \sum_z a(z) S(z) \ket{z}\bra{z},
&
\Avg{A,B}_\sigma &= \sum_z a(z) b(z) S(z),
\\
\mc F^*(t) A &= \sum_{z,z'} a(z') P_{Z(t)|Z(0)}(z'|z) \ket{z}\bra{z},
&
\mc F_*(t) A &= 
\sum_{z,z'} \frac{P_{Z(t)|Z(0)}(z|z') S(z')}{S(z)}  a(z')\ket{z}\bra{z},
\\
\mc L^* A &= \sum_{z,z'} a(z') r(z'|z) \ket{z}\bra{z},
&
\mc L_* A &= \sum_{z,z'} \frac{r(z|z') S(z') }{S(z)} a(z')\ket{z}\bra{z}.
\end{align}
$\mc F^*(t)$ corresponds to the predictive conditional expectation,
while $\mc F_*(t)$ corresponds to the retrodictive conditional
expectation via Bayes theorem.

The quantum detailed-balance condition given by Eqs.~(\ref{DB2}) can
be applied to the classical case by assuming
\begin{align}
\Theta(\ket{z}\bra{z}) &= \ket{T^{-1} z}\bra{T^{-1} z},
\end{align}
where $z$ is assumed to be a column vector and $T$ is some invertible
matrix. Then Eqs.~(\ref{DB2}) imply
\begin{align}
r(z'|z) S(z) &= r(T z|T z') S(T z'),
&
S(z) &= S(T z),
\label{CDB}
\end{align}
which agree with the classical definition \cite[Eqs.~(5.3.45) and
(5.3.46)]{gardiner}.

If the parametrization of the initial state is expressed in terms of a
parametrized distribution $P(z|\theta)$ as
\begin{align}
\rho(0) &= \tau(\theta) = \sum_z P(z|\theta) \ket{z}\bra{z},
\end{align}
then
\begin{align}
P(z|0) &= S(z),
\\
\mc F(t)\tau(\theta) &= \sum_z P(z,t|\theta) \ket{z}\bra{z},
\\
P(z,t|\theta) &\equiv \sum_{z'} P_{Z(t)|Z(0)}(z|z') P(z'|\theta),
\\
J_{jk}(t) &= \sum_z \frac{[\partial_j P(z,t|\theta)][\partial_k P(z,t|\theta)]}{S(z)}
= \sum_z S(z) s_j(z,t)s_k(z,t),
\\
s_j(z,t) &\equiv \frac{1}{S(z)}\partial_j P(z,t|\theta)
= \partial_j \ln P(z,t|\theta)
=\sum_{z'}  \frac{P_{Z(t)|Z(0)}(z|z') S(z')}{S(z)} s_j(z',0),
\\
X_j(t) &= \sum_z s_j(z,t)\ket{z}\bra{z},
\\
\dot s_j(z,t) &= \sum_{z'} \frac{r(z|z') S(z')}{S(z)} s_j(z',t).
\end{align}
$J$ coincides with the classical Fisher information, while $s_j$
coincides with the classical score function.

\section{\label{app_gauss}Gaussian states and channels}
This appendix follows Refs.~\cite{holevo11,holevo19}. Let $Q$ be the
vector of canonical observables of $s$ bosonic modes given by
Eq.~(\ref{Q}) and $W(z)$ be the Weyl operator defined by
\begin{align}
W(z) &\equiv \exp(i z^jQ_j),
\\
z &\equiv \mqty(x^1 & \dots & x^s & y^1 & \dots & y^s) \in \mb R^{2s}.
\end{align}
Some useful identities are as follows.
\begin{align}
\frac{1}{2} \Bk{ W(z) Q_j +Q_j W(z)} &= -i \pdv{}{z^j} W(z),
\label{QW}
\\
[W(z),Q_j] &= \Omega_{jk} z^k W(z).
\label{WQ}
\end{align}
A Gaussian state $\rho$ is defined by having a Gaussian characteristic
function in the form of
\begin{align}
  \trace \rho W(z) &= \exp\bk{i z^j m_j  - \frac{1}{2} z^j\Sigma_{jk} z^k},
  \\
  m_j &\equiv \trace \rho Q_j,
  \\
  \Sigma_{jk} &\equiv \Avg{Q_j-m_j,Q_k-m_k}_{\rho},
\end{align}
where $m$ is the mean vector, $\Sigma$ is the covariance matrix, and
the weighted inner product $\avg{\cdot,\cdot}_{\rho}$ here
is defined with respect to the Helstrom density map
$\mc E_{\rho,\Hel}$ given by Eq.~(\ref{Hel}). Such a Gaussian
state is labeled by $(m,\Sigma)$. 

\begin{proposition}[{Ref.~\cite[Proposition~5.6.3]{holevo11}}]
\label{prop_sld}
For an $(m,\Sigma)$ Gaussian state $\rho$ and the Helstrom density map
given by Eq.~(\ref{Hel}),
\begin{align}
i [Q_j,\rho] &= \Omega_{jl}\Sigma^{lk}\mc E_{\rho,\Hel} (Q_k- m_k),
\label{score_unitary}
\end{align}
where $\Sigma$ with the superscripts denotes the inverse of $\Sigma$, viz.,
\begin{align}
\Sigma_{jl}\Sigma^{lk} &= \delta_j^k,
\end{align}
and $\delta_j^k$ is the Kronecker delta.
\end{proposition}
\begin{proof}
  Using Eq.~(\ref{WQ}), the characteristic function of the left-hand
  side of Eq.~(\ref{score_unitary}) becomes
\begin{align}
i  \trace W(z) [Q_j,\rho]
&= i \trace [W(z),Q_j] \rho
= i \Omega_{jl}z^l  \trace W(z) \rho.
\label{lhs}
\end{align}
while the characteristic function of the right-hand
side of Eq.~(\ref{score_unitary}),
using Eq.~(\ref{QW}), becomes
\begin{align}
\Omega_{jl} \Sigma^{lk}\trace W(z) \mc E_{\rho,\Hel} (Q_k - m_k)
&= \Omega_{jl} \Sigma^{lk}\bk{-i \pdv{}{z^k}- m_k} \trace W(z) \rho
= i\Omega_{jl}  z^l \trace W(z) \rho,
\end{align}
which is equal to Eq.~(\ref{lhs}).
\end{proof}

A Gaussian channel $\mc F$ that maps $s$ modes to $s'$ modes
is defined by
\begin{align}
\mc F^* \tilde W(\zeta) &= g(\zeta) W(\zeta F),
\label{pull_W}
\\
(\zeta F)^j &= \zeta^\mu F_\mu{}^j ,
\\
g(\zeta) &= \exp\bk{i \zeta^\mu l_\mu  - \frac{1}{2} \zeta^\mu S_{\mu\nu} \zeta^\nu},
\label{g}
\end{align}
where $\tilde W(\zeta)$ is the Weyl operator for $s'$ modes,
$\zeta \in \mb R^{2s'}$, $l \in \mb R^{2s'}$,
$F \in \mb R^{2s'\times 2s}$, and $S \in \mb R^{2s'\times 2s'}$. Such
a Gaussian channel is labeled by $(l,F,S)$. It follows that the
characteristic function after an application of an $(l,F,S)$ channel
to an $(m,\Sigma)$ state is given by
\begin{align}
\trace (\mc F\rho) \tilde W(\zeta) &= \trace \rho \mc F^* \tilde W(\zeta)
=\exp\bk{i \zeta^\mu \tilde m_\mu - \frac{1}{2} \zeta^\mu \tilde\Sigma_{\mu\nu} \zeta^\nu},
\\
\tilde m_\mu &= F_\mu{}^j m_j  + l_\mu,
\label{m2}
\\
\tilde\Sigma_{\mu\nu} &= F_\mu{}^j \Sigma_{jk} F_\nu{}^k + S_{\mu\nu}.
\label{Sigma2}
\end{align}
In matrix form, the last two equations can be expressed as
\begin{align}
\tilde m &= F m + l,
&
\tilde\Sigma &= F \Sigma F^\top + S.
\end{align}
\begin{proposition}
\label{prop_gce}
For an $(m,\Sigma)$ Gaussian state, an $(l,F,S)$ Gaussian channel
$\mc F$, and the Helstrom density map given by Eq.~(\ref{Hel}),
\begin{align}
\mc F^* Q_\mu &= F_\mu{}^j Q_j + l_\mu,
\label{pull_Q}
\\
\mc F_* Q_j &= m_j + \tilde F_j{}^\mu \bk{Q_\mu - \tilde m_\mu},
\\
\tilde F_j{}^\mu &\equiv \Sigma_{jk} F_\nu{}^k \tilde\Sigma^{\nu\mu},
\label{push_Q} 
\end{align}
where $\tilde m$ is given by Eq.~(\ref{m2}), $\tilde\Sigma$ is given
by Eq.~(\ref{Sigma2}), and $\tilde\Sigma$ with the superscripts
denotes the inverse of $\tilde\Sigma$. Eq.~(\ref{push_Q})
can also be expressed in the matrix form
\begin{align}
\tilde F &= \Sigma F^\top \tilde\Sigma^{-1}.
\end{align}
\end{proposition}
\begin{proof}
  Eq.~(\ref{push_Q}) has been derived in
  Ref.~\cite[Proposition~1]{gce2}.  To derive Eq.~(\ref{pull_Q}), use
  Eq.~(\ref{QW}) and Eqs.~(\ref{pull_W})--(\ref{g}) to write
\begin{align}
\mc F^* Q_\mu
&= \mc F^* [Q_\mu \circ \tilde W(0)]
= \left.-i \pdv{}{\zeta^\mu} \mc F^*\tilde W(\zeta)\right|_{\zeta=0}
= \left.-i \pdv{}{\zeta^\mu}  g(\zeta) W(\zeta F)\right|_{\zeta = 0}
= l_\mu + F_\mu{}^j Q_j.
\end{align}
\end{proof}

\section{\label{app_thermal} Some results for coupled and damped
  harmonic oscillators}
\begin{proposition}\label{prop_thermal}
  For the thermal state $\sigma$ given by Eq.~(\ref{thermal}) and the
  generator $\mc L$ given by Eq.~(\ref{GKSL}),
\begin{align}
\mc L\sigma = 0.
\label{Lsigma}
\end{align}
\end{proposition}
\begin{proof}
  Using the basic identities $a_j\ket{n_j} = \sqrt{n_j} \ket{n_j-1}$
  and $a_j^\dagger \ket{n_j} = \sqrt{n_j+1} \ket{n_j+1}$ for a
  single-mode Fock state $\ket{n_j}$ and expressing $\sigma$ in terms
  of the Fock basis, it can be shown that
\begin{align}
a_j \sigma &= z \sigma a_j,
&
a_j^\dagger \sigma &= z^{-1} \sigma a_j^\dagger,
&
z &\equiv \frac{\bar n}{\bar n+1}.
\label{a_sigma}
\end{align}
It follows that
\begin{align}
a_j^\dagger a_k \sigma &= \sigma a_j^\dagger a_k,
&
a_k \sigma a_j^\dagger &= z \sigma a_k a_j^\dagger,
&
a_j^\dagger \sigma a_k &= z^{-1} \sigma a_j^\dagger a_k.
\end{align}
Plugging these expressions into Eq.~(\ref{GKSL}) for $\rho = \sigma$
leads to Eq.~(\ref{Lsigma}).

\end{proof}

The next proposition is a general condition for the weighted adjoints
to be the same for any density map in Petz's class.
\begin{proposition}
\label{prop_Delta}
Given a map $\mc M^*:\mc O(\mc H)\to\mc O(\mc H)$, define
$\mc M_*:\mc O(\mc H)\to\mc O(\mc H)$ as the adjoint of $\mc M^*$ with
respect to the weighted inner product given by Eq.~(\ref{weighted}),
viz.,
\begin{align}
\Avg{A,\mc M_* B}_\sigma &= \Avg{\mc M^*A, B}_\sigma.
\label{M_def}
\end{align}
If $\mc M^*$ commutes with the $\Delta_\sigma$ map defined by
Eq.~(\ref{Delta}), then
\begin{align}
\mc M_* &= \mc M_{*,\mc R}
\end{align}
for any density map $\mc E_\sigma$ in the form of Eq.~(\ref{petz}),
where $\mc M_{*,\mc R}$ is the weighted adjoint of $\mc M^*$
in terms of the right product map $\mc E_\sigma = \mc R_\sigma$
defined by Eq.~(\ref{R_map}). To be explicit,
\begin{align}
\Avg{A,\mc R_\sigma \mc M_{*,\mc R}  B} &= \Avg{\mc M^* A,\mc R_\sigma B}
\quad
\forall A,B \in \mc O(\mc H),
\label{Mr}
\\
\mc M_{*,\mc R} &= \mc R_\sigma^{-1} \mc M \mc R_\sigma.
\end{align}
\end{proposition}
\begin{proof}
For any $A,B \in \mc O(\mc H)$,
\begin{align}
\Avg{ A,\mc M_*B}_\sigma &= \Avg{\mc M^* A, B}_\sigma
&
(\textrm{by Eq.~(\ref{M_def})})
\\
&= \Avg{\mc R_\sigma \phi(\Delta_\sigma) \mc M^* A, B}
&
(\textrm{by Eqs.~(\ref{weighted}) and (\ref{petz}),
$\mc E_\sigma = \mc E_\sigma^*$})
\\
&= \Avg{\mc M^* \phi(\Delta_\sigma) A, \mc R_\sigma  B}
&
(\textrm{$\mc M^*$ and $\Delta_\sigma$ commute,
$\mc R_\sigma = \mc R_\sigma^*$})
\\
&= \Avg{\phi(\Delta_\sigma) A,\mc R_\sigma \mc M_{*,\mc R}   B}
&
(\textrm{by Eq.~(\ref{Mr})})
\\
&= \Avg{\mc R_\sigma\phi(\Delta_\sigma) A,\mc M_{*,\mc R} B}
&
(\mc R_\sigma = \mc R_\sigma^*)
\\
&= \Avg{A,\mc M_{*,\mc R} B}_\sigma.
&
(\textrm{by Eqs.~(\ref{weighted}) and (\ref{petz}))})
\end{align}
\end{proof}
\begin{proposition}
\label{prop_commute}
For the thermal steady state given by Eq.~(\ref{thermal}) and the
semigroup generator given by Eq.~(\ref{GKSL}), $\mc L^*$ commutes with
$\Delta_\sigma$. Hence, $\mc L_*$ is the same for all density maps in
Petz's class by Prop.~\ref{prop_Delta}.
\end{proposition}
\begin{proof}
Given Eq.~(\ref{GKSL}), $\mc L^*$ can be expressed as
\begin{align}
\mc L^* A &= -i \omega_{jk} \Bk{a_j^\dagger a_k,A}
+ \frac{1}{2}(\bar n+1)\gamma_{jk}  (2 a_j^\dagger A a_k
- A a_j^\dagger a_k  -  a_j^\dagger a_k A)
+ \frac{1}{2}\bar n\gamma_{jk}
(2 a_k A a_j^\dagger   - A a_k a_j^\dagger  - a_k a_j^\dagger A)
\label{GKSL2}
\end{align}
for any $A \in \mc O(\mc H)$. Eqs.~(\ref{a_sigma})  imply
\begin{align}
\sigma^{-1} a_j &= z a_j \sigma^{-1},
&
a_j^\dagger \sigma^{-1} &= z \sigma^{-1} a_j^\dagger.
\label{a_sigma2}
\end{align}
It follows from Eqs.~(\ref{a_sigma}) and (\ref{a_sigma2}) that
\begin{align}
a_j^\dagger a_k (\Delta_\sigma A) &= \Delta_\sigma (a_j^\dagger a_k A),
&
(\Delta_\sigma A) a_j^\dagger a_k &= \Delta_\sigma (A a_j^\dagger a_k),
\\
a_j^\dagger (\Delta_\sigma A) a_k &= 
\Delta_\sigma (a_j^\dagger A a_k),
&
a_k (\Delta_\sigma A) a_j^\dagger &= \Delta_\sigma (a_k A a_j^\dagger).
\end{align}
Hence, all the operations on $A$ on the right-hand side of
Eq.~(\ref{GKSL2}) commute with $\Delta_\sigma$, giving
$\mc L^* \Delta_\sigma = \Delta_\sigma \mc L^*$.
\end{proof}

\section{\label{app_reverse}Time reversal on a larger Hilbert space}
Given a CPTP map $\mc F$, it can always be represented in a larger
Hilbert space as
\begin{align}
\mc F \rho &= \trace' U (\rho\otimes\chi) U^\dagger,
\label{F2}
\end{align}
where $\chi$ is a density operator on a reservoir Hilbert space
$\mc H'$, $U$ is a unitary operator on $\mc H\otimes\mc H'$, and
$\trace'$ denotes the partial trace with respect to $\mc H'$. Now
define another unitary operator $V$ on $\mc H\otimes \mc H'$ by
\begin{align}
V &\equiv \Theta U^\dagger
\label{UV}
\end{align}
with respect to an antiunitary map $\Theta$ on
$\mc O(\mc H\otimes \mc H')$. If $U$ is expressed as
$U = \exp(-iHt)$ with respect to a Hamiltonian $H$, then
$V = \exp(-i\Theta H t)$. Assume that $\Theta$ is separable in the
sense of
\begin{align}
\Theta (A\otimes B) &= (\Theta A) \otimes (\Theta B),
\end{align}
and suppose that another CPTP map $\mc G$ is given by
\begin{align}
\mc G \rho &= \trace' V (\rho\otimes\Theta\chi) V^\dagger.
\label{G2}
\end{align}
The following proposition relating $\mc F$ and $\mc G$ is a
generalization of Ref.~\cite{carmichael76}, which focuses on the
$\mc E_\sigma = \mc R_\sigma$ case.
\begin{proposition}
\label{prop_reverse}
Assume that $\sigma\otimes\chi$ is a steady state of the total system, viz.,
\begin{align}
U(\sigma\otimes\chi) U^\dagger &= \sigma\otimes\chi.
\label{steady_total}
\end{align}
Then the $\mc F$ and $\mc G$ maps given by Eqs.~(\ref{F2})--(\ref{G2})
satisfy the reversal relation given by Eq.~(\ref{reverse_map}).
\end{proposition}
The proof is delegated to Appendix~\ref{app_math}.

Eq.~(\ref{steady_total}) is an additional assumption about the steady
state of the total system. It implies that $\sigma$ is a steady state
of $\mc F$, but the converse need not be true. This assumption was
also made by Carmichael and Walls---who called it microscopic
stationarity \cite[Eq.~(2.9)]{carmichael76}---to derive Agarwal's
detailed balance from time-reversal symmetry.

For a system satisfying time-reversal symmetry, $U = V$.  By virtue of
Prop.~\ref{prop_reverse}, the time-reversal symmetry together with the
assumption of a $\Theta$-symmetric steady state
$\Theta(\sigma\otimes\chi) = \sigma\otimes\chi$ lead to
$\mc G = \mc F$ and the detailed-balance condition defined by
Eqs.~(\ref{DB2}).

\section{\label{app_crooks}Another definition of time reversal}
Let $\mc A,\mc B:\mc O(\mc H)\to \mc O(\mc H)$ be two linear maps and
denote a certain dual of $\mc A$ by
$\mc A^\#:\mc O(\mc H)\to \mc O(\mc H)$.  A criterion of the dual
proposed by Crooks \cite{crooks08} and generalized by Manzano
\textit{et al.}\ \cite{manzano15,manzano18} is
\begin{align}
\trace \mc B \mc A \sigma &= \trace \mc A^\# \mc B^\# \Theta\sigma,
\label{dual}
\end{align}
where $\Theta \sigma \equiv \theta \sigma \theta^{-1}$ and $\theta$ is
a unitary or antiunitary operator. Here I focus on the antiunitary
case.  It turns out that their assumed form of $\mc A^\#$ is only one
possible solution of Eq.~(\ref{dual}); Eq.~(\ref{reverse_map}) for a
class of density maps can satisfy it as well.
\begin{proposition}
\label{prop_crooks}
Suppose that $\mc E_\sigma$ is self-adjoint with respect to the
Hilbert-Schmidt inner product ($\mc E_\sigma = \mc E_\sigma^*$) and
satisfies $\mc E_\sigma I = \sigma$ and $\mc E_\sigma^{-1}\sigma =
I$. In particular, any normalized density map in Petz's class
suffices. If one sets
\begin{align}
\mc A^\# &= (\Theta \mc A_* \Theta^*)^* = \Theta \mc E_\sigma \mc A^* \mc E_\sigma^{-1}\Theta^*,
\label{dual2}
\end{align}
then Eq.~(\ref{dual}) is satisfied.
\end{proposition}
\begin{proof}
\begin{align}
\trace \mc A^\# \mc B^\# \Theta\sigma &= 
\trace \Theta\mc E_\sigma \mc A^* \mc E_\sigma^{-1}
 \Theta^* \Theta \mc E_\sigma \mc B^* \mc E_\sigma^{-1} \Theta^* \Theta \sigma
&
(\textrm{by Eq.~(\ref{dual2})})
\\
&= \trace \Theta\mc E_\sigma \mc A^* \mc B^* \mc E_\sigma^{-1} \sigma
&
(\Theta^* = \Theta^{-1})
\\
&= \trace \Theta \mc E_\sigma \mc A^* \mc B^* I
&
(\mc E_\sigma^{-1} \sigma = I)
\\
&= \Avg{I,\Theta \mc E_\sigma \mc A^* \mc B^* I}
&
(\textrm{by Eq.~(\ref{HS})})
\\
&= \Avg{\mc A^* \mc B^* I,\mc E_\sigma \Theta^* I}
&
(\textrm{by definition of $\Theta^*$, $\mc E_\sigma = \mc E_\sigma^*$})
\\
&= \Avg{\mc A^* \mc B^* I,\sigma}
&
(\Theta^*I = I, \mc E_\sigma I = \sigma)
\\
&= \Avg{I,\mc B \mc A \sigma}
&
(\textrm{by definition of Hilbert-Schmidt adjoint})
\\
&= \trace \mc B \mc A \sigma.
&
(\textrm{by Eq.~(\ref{HS})})
\end{align}
\end{proof}
Eq.~(\ref{reverse_map}) can be written as $\mc G = \mc F^\#$ given
Eq.~(\ref{dual2}).

Suppose that $\mc A$ can be expressed in terms of some 
operators $\{\alpha_j\in\mc O(\mc H)\}$ as
\begin{align}
\mc A \rho &= \sum_j \alpha_j \rho \alpha_j^\dagger.
\end{align}
Assume also the density map $\mc E_{\sigma,\Con}$ given by
Eq.~(\ref{connes}).  Then Eq.~(\ref{dual}) can be written as
\begin{align}
\mc A^\# \rho
&= \sum_j \tilde\alpha_j \rho \tilde\alpha_j^\dagger,
&
\tilde\alpha_j &\equiv \Theta \Delta_\sigma^{1/2} \alpha_j^\dagger
= \theta \sigma^{1/2} \alpha_j^\dagger \sigma^{-1/2}  \theta^{-1}.
\end{align}
This form of $\mc A^\#$ is assumed by Crooks and Manzano \textit{et
  al.}\ \cite{crooks08,manzano15,manzano18}, but it is not the only
form that satisfies Eq.~(\ref{dual}), as Prop.~\ref{prop_crooks}
demonstrates.

\section{\label{app_math}Miscellaneous results and proofs}

\begin{proof}[Proof of Theorem~\ref{thm}]
  The most nontrivial part of the theorem is the relation between the
  definition of the geometric variable given by Eq.~(\ref{x}) and the
  equation of motion given by Eq.~(\ref{xrate}). To prove that, I
  first derive Eq.~(\ref{K}) as follows:
\begin{align}
\dot J_{jk}(t) &= 
\Avg{\dot X_j(t),X_k(t)}_\sigma + \Avg{X_j(t),\dot X_k(t)}_\sigma 
&
(\textrm{by Eq.~(\ref{J2})})
\\
&= \Avg{\mc L_* X_j(t),X_k(t)}_\sigma + \Avg{X_j(t),\mc L_* X_k(t)}_\sigma
&
(\textrm{by Eq.~(\ref{Xrate})})
\\
&= \Avg{X_j(t),(\mc L^*+\mc L_*) X_k(t)}_\sigma.
&
(\textrm{by Eq.~(\ref{adjoint2})})
\end{align}
Now I can go from Eq.~(\ref{x}) to Eq.~(\ref{xrate}) as follows:
\begin{align}
\dot x_j(t) &= \trace\Bk{\dot X_j(t) \rho (t) + X_j(t)\dot\rho(t)}
&
(\textrm{by Eq.~(\ref{x})})
\label{xrate2}
\\
&= \trace\Bk{\dot X_j(t)\rho(t) + X_j(t) \mc L \rho(t)}
&
(\textrm{by Eqs.~(\ref{F})})
\\
&= \trace\BK{\dot X_j(t)\sigma
+ \epsilon v^k
\Bk{\dot X_j(t)\mc F(t)\partial_k\tau + 
X_j(t) \mc L \mc F(t)\partial_k  \tau}} + o(\epsilon)
&
(\textrm{by Eqs.~(\ref{steady}) and (\ref{taylor})})
\\
&=  \epsilon v^k\trace\BK{\Bk{\mc L_* X_j(t)} \mc E_\sigma X_k(t) + 
X_j(t) \mc L \mc E_\sigma X_k(t)} + o(\epsilon)
&
(\textrm{by Eqs.~(\ref{score}), (\ref{Xmean}), and (\ref{Xrate})})
\\
&= \epsilon v^k\Bk{\Avg{\mc L_* X_j(t),X_k(t)}_\sigma
+ \Avg{X_j(t),\mc L_* X_k(t)}_\sigma} + o(\epsilon)
&
(\textrm{by Eqs.~(\ref{weighted}) and (\ref{L_push})})
\\
&= -\epsilon v^k K_{jk}(t) + o(\epsilon).
&
(\textrm{by Eqs.~(\ref{adjoint2}) and (\ref{K})})
\end{align}
Eq.~(\ref{R2}) is obtained simply by combining Eqs.~(\ref{R}),
(\ref{f}), (\ref{R2}), and (\ref{xrate}).  $K(t)$ is symmetric because
$J(t)$ is symmetric as per Eq.~(\ref{Jsym}).  $K(t)$ is
positive-semidefinite because $J$ is monotonic as per
Eq.~(\ref{J_mono}).
\end{proof}

\begin{proof}[Proof of Theorem~\ref{thm2}]
Following steps similar to the proof of Theorem~\ref{thm},
it is straightforward to show that
\begin{align}
\dot y_j(t) &= -\Avg{X_j(0),\mc L_* X_k(t)}_\sigma f^k + o(\epsilon).
\label{yrate2}
\end{align}
This current as a function of $t$ does not seem to have any simple
relation with the other quantities, but a relation at $t = 0$ is
possible.  Eqs.~(\ref{yrate}) and (\ref{O}) are obtained by plugging
$t = 0$ in Eq.~(\ref{yrate2}). Eq.~(\ref{KO}) follows from
Eq.~(\ref{K}) in Theorem~\ref{thm}.  $O$ is positive-semidefinite
because
\begin{align}
v^jO_{jk} v^k &= \frac{1}{2} v^j K_{jk}(0) v^k 
\quad\forall v
\label{OK}
\end{align}
and $K$ is positive-semidefinite as per
Theorem~\ref{thm}. Eq.~(\ref{R3}) follows from Eqs.~(\ref{R2}),
(\ref{yrate}), and (\ref{OK}).
\end{proof}

Before proving Theorems~\ref{thm3} and \ref{thm4} and
Props.~\ref{prop_reverse} and \ref{prop_reverse_map}, I need the
following lemma.
\begin{lemma}
\label{lem_petz}
Assume a density map $\mc E_\rho$ in the form of Eq.~(\ref{petz}).
\begin{enumerate}
\item For any unitary or  antiunitary map $\mc U$,
\begin{align}
\mc U \mc E_\rho  &= \mc E_{\mc U\rho} \mc U,
\label{UE}
\\
\mc U \mc E_\rho^{-1}  &= \mc E_{\mc U\rho}^{-1} \mc U.
\label{UE2}
\end{align}
\item For any $A \in \mc O(\mc H)$, $\rho \in \mc P(\mc H)$,
  $\chi \in \mc P(\mc H')$, and $B \in \mc O(\mc H')$ that commutes
  with $\chi$,
\begin{align}
\mc E_{\rho\otimes\chi} (A\otimes B) &= (\mc E_\rho A) \otimes (B\chi).
\label{E_tensor}
\end{align}
\end{enumerate}
\end{lemma}
\begin{proof}
  Assume the Hilbert-Schmidt inner product in the following.  The
  $\Delta_\rho$ map defined by Eq.~(\ref{Delta}) is self-adjoint and
  positive-definite, because
\begin{align}
\Avg{A,\Delta_\rho A} &= \trace A^\dagger \rho A \rho^{-1}
= \trace (\rho A \rho^{-1})^\dagger A = \Avg{\Delta_\rho A,A},
\\
\Avg{A,\Delta_\rho A} &= \trace \rho^{-1/2}
A^\dagger \rho^{1/2} \rho^{1/2} A \rho^{-1/2}
= \Avg{\rho^{1/2} A \rho^{-1/2},\rho^{1/2} A \rho^{-1/2}} \ge 0,
\end{align}
and $\Delta_\rho^{-1} A = \rho^{-1} A \rho$ exists.  Now define
$\{E_j \in \mc O(\mc H)\}$ as the orthonormal eigenvectors of
$\Delta_\rho$ and $\{\lambda_j\}$ as the eigenvalues, such that
\begin{align}
\Delta_\rho E_j &= \lambda_j E_j,
&
\Avg{E_j,E_k} &= \delta_{jk},
\end{align}
where Einstein summation is abandoned for clarity. 
Notice that
\begin{align}
\mc U \Delta_\rho A &= (\mc U \rho) (\mc U A) (\mc U \rho)^{-1}
= \Delta_{\mc U \rho} \mc U A,
\end{align}
so if $E_j$ is an eigenvector of $\Delta_\rho$ with eigenvalue
$\lambda_j$, then $\mc U E_j$ is an eigenvector of
$\Delta_{\mc U\rho}$ with the same eigenvalue $\lambda_j$.
$\mc E_\rho$ and $\mc E_{\mc U\rho}$ can then be expressed as
\begin{align}
\mc E_\rho A &=  \mc R_\rho \phi(\Delta_\rho) A
= \sum_j \phi(\lambda_j) E_j \rho \Avg{E_j,A},
\\
\mc E_{\mc U\rho} A &= 
\mc R_{\mc U\rho} \phi(\Delta_{\mc U\rho}) A
= \sum_j \phi(\lambda_j) (\mc U E_j)(\mc U\rho) \Avg{\mc U E_j,A}.
\label{EU}
\end{align}
If $\mc U$ is unitary, 
\begin{align}
\mc U \mc E_\rho A &=  \sum_j \phi(\lambda_j) 
(\mc U E_j) (\mc U\rho) \Avg{E_j,A}
&
(\mc U(AB) = (\mc U A) (\mc U B))
\\
&= \sum_j \phi(\lambda_j) 
(\mc U E_j) (\mc U\rho) \Avg{\mc U E_j,\mc U A}
&
(\textrm{$\mc U$ is unitary})
\\
&= \mc E_{\mc U\rho} \mc U A.
&
(\textrm{by Eq.~(\ref{EU})})
\end{align}
Similarly, if $\mc U$ is antiunitary,
\begin{align}
\mc U \mc E_\rho A &=  \sum_j \phi(\lambda_j) 
(\mc U E_j) (\mc U\rho) \Avg{A,E_j}
&
(\mc U(AB) = (\mc U A) (\mc U B))
\\
&= \sum_j \phi(\lambda_j) 
(\mc U E_j) (\mc U\rho) \Avg{\mc U E_j,\mc U A}
&
(\textrm{$\mc U$ is antiunitary})
\\
&= \mc E_{\mc U\rho} \mc U A.
&
(\textrm{by Eq.~(\ref{EU})})
\end{align}
Hence Eq.~(\ref{UE}) is proved. To prove Eq.~(\ref{UE2}), use
Eq.~(\ref{UE}) and $\mc U^{-1} = \mc U^*$ to obtain
$\mc E_{\mc U\rho} = \mc U \mc E_\rho \mc U^*$ and
$\mc E_{\mc U\rho}^{-1} = \mc U \mc E_\rho^{-1} \mc U^*$.  To prove
Eq.~(\ref{E_tensor}), assume any $B$ that commutes with $\chi$ and
note that
\begin{align}
\Delta_{\rho\otimes\chi} (A\otimes B) &= (\Delta_\rho A) \otimes B.
\end{align}
It follows that, if $E_j$ is an eigenvector of $\Delta_\rho$ with
eigenvalue $\lambda_j$, $E_j\otimes B$ is an eigenvector of
$\Delta_{\rho\otimes\chi}$ with the same eigenvalue. Writing
$A = \sum_j E_j \avg{E_j,A}$, one obtains
\begin{align}
\mc E_{\rho\otimes\chi} (A\otimes B)
&= \mc R_{\rho\otimes\chi}\phi(\Delta_{\rho\otimes\chi}) (A\otimes B) 
= \sum_j \phi(\lambda_j) (E_j\rho \otimes B \chi) \Avg{E_j,A} 
= (\mc E_\rho A) \otimes (B \chi).
\end{align}
\end{proof}

\begin{proof}[Proof of Theorem~\ref{thm3}]
\begin{align}
-\tilde O_{jk} &= \Avg{\tilde X_j(0),\mc K_* \tilde X_k(0)}_{\tilde\sigma}
&
(\textrm{by Eq.~(\ref{O2})})
\\
&=\Avg{\mc K^* \Theta X_j(0),\Theta X_k(0)}_{\tilde\sigma}
&
(\textrm{by definition of $\mc K_*$ and Eq.~(\ref{X2})})
\\
&= \Avg{\Theta \mc L_* X_j(0),\Theta X_k(0)}_{\tilde\sigma}
&
(\textrm{by Eq.~(\ref{reverse_gen})})
\\
&= \Avg{\Theta \mc L_* X_j(0),\mc E_{\Theta\sigma} \Theta X_k(0)}
&
(\textrm{by Eq.~(\ref{weighted}) and (\ref{steady2})})
\\
&= \Avg{\Theta \mc L_* X_j(0),\Theta \mc E_{\sigma} X_k(0)}
&
(\textrm{by Lemma~\ref{lem_petz}})
\\
&= \Avg{\mc E_{\sigma} X_k(0),\mc L_* X_j(0)}
&
(\textrm{$\Theta$ is antiunitary})
\\
&= -O_{kj}.
&
(\textrm{by Eq.~(\ref{O}), $\mc E_\sigma = \mc E_\sigma^*$})
\end{align}
\end{proof}

\begin{proof}[Proof of Theorem~\ref{thm4}]
\begin{align}
-O_{jk} &= \Avg{X_j(0), \mc L_* X_k(0)}_{\sigma}
&
(\textrm{by Eq.~(\ref{O})})
\\
&= \Avg{X_j(0), \Theta^*\mc K^* \Theta X_k(0)}_{\sigma}
&
(\textrm{by Eq.~(\ref{reverse_gen}) and $\Theta^* = \Theta^{-1}$})
\\
&= \Avg{\mc K^* \Theta X_k(0),\Theta X_j(0)}_{\Theta\sigma}
&
(\textrm{by definition of $\Theta^*$ and Lemma~\ref{lem_petz}})
\\
&= T_k{}^lT_j{}^m\Avg{\mc K^* X_l(0),X_m(0)}_{\sigma}
&
(\textrm{by Eqs.~(\ref{identical}) and (\ref{T})})
\\
&= -T_k{}^lT_j{}^m \tilde O_{lm}.
&
(\textrm{by Eqs.~(\ref{O2}) and (\ref{identical})})
\end{align}
\end{proof}

\begin{proof}[Proof of Proposition~\ref{prop_reverse}]
  Define 
\begin{align}
\mc U A &\equiv U A U^\dagger,
\label{U2}
\\
\mc V A &\equiv V A V^\dagger,
\end{align}
which are unitary maps with respect to the Hilbert-Schmidt inner
product. Eq.~(\ref{UV}) leads to
\begin{align}
\mc V &= \Theta \mc U^* \Theta^*.
\label{UV2}
\end{align}
Then one can write, for any $A,B \in \mc O(\mc H)$,
\begin{align}
\Avg{A,\mc F_* B}_\sigma &= \Avg{A,\mc F \mc E_\sigma B}
\label{AFB}
&
(\textrm{by Eqs.~(\ref{weighted}) and (\ref{retro})})
\\
&= \Avg{A,\trace' \mc U (\mc E_\sigma B \otimes \chi)}
&
(\textrm{by Eqs.~(\ref{F2}) and (\ref{U2})})
\\
&= \Avg{A\otimes I, \mc U\mc E_{\sigma\otimes\chi}(B\otimes I)}
&
(\textrm{by definition of $\trace'$ and Lemma~\ref{lem_petz}})
\\
&=\Avg{\mc U^*(A\otimes I), \mc E_{\sigma\otimes\chi}(B\otimes I)}.
&
(\textrm{by definition of $\mc U^{*}$})
\label{Fchurch}
\end{align}
Similarly, one can write
\begin{align}
\Avg{A,\Theta^* \mc G^* \Theta B}_\sigma
\label{ATGTB}
&= \Avg{\mc E_\sigma A,\Theta^* \mc G^* \Theta B}
&
(\textrm{by Eq.~(\ref{weighted}), $\mc E_\sigma = \mc E_\sigma^*$})
\\
&= \Avg{\mc G^* \Theta B,\mc E_{\Theta \sigma}\Theta A}
&
(\textrm{by definition of $\Theta^*$, Lemma~\ref{lem_petz}})
\\
&= 
\Avg{\mc V^*\Theta(B\otimes I),\mc E_{\Theta\sigma\otimes\Theta\chi} \Theta (A\otimes I)},
\label{Gchurch}
\end{align}
where the last step follows the same arguments that give
Eq.~(\ref{Fchurch}).  Now Eqs.~(\ref{Fchurch}) and (\ref{Gchurch}) can
be shown to be equal as follows:
\begin{align}
\Avg{\mc V^*\Theta(B\otimes I),\mc E_{\Theta\sigma\otimes\Theta\chi} \Theta (A\otimes I)}
&= 
\Avg{\Theta^*\mc V\mc E_{\Theta\sigma\otimes\Theta\chi} \Theta (A\otimes I),
B\otimes I}
&
(\textrm{by definition of $\mc V^*$ and $\Theta^*$})
\\
&= \Avg{\mc U^* \Theta^* \mc E_{\Theta\sigma\otimes\Theta\chi} \Theta (A\otimes I),
B\otimes I}
&
(\textrm{by Eq.~(\ref{UV2})})
\\
&= \Avg{\mc E_{\mc U^*(\sigma\otimes\chi)} \mc U^* (A\otimes I),B\otimes I}
&
(\textrm{by Lemma~\ref{lem_petz}, $\Theta^* = \Theta^{-1}$})
\\
&= \Avg{\mc E_{\sigma\otimes\chi} \mc U^* (A\otimes I),B\otimes I}.
&
(\textrm{by Eq.~(\ref{steady_total})})
\end{align}
Hence, the left-hand sides of Eqs.~(\ref{AFB}) and (\ref{ATGTB}) are
equal, leading to Eq.~(\ref{reverse_map}).
\end{proof}

\begin{proposition}
\label{prop_reverse_map}
With the reversal relation given by Eq.~(\ref{reverse_map}) for the
$\mc F$ and $\mc G$ maps and the relation for their steady states
given by Eq.~(\ref{steady2}), $\mc F$ is also the reverse of $\mc G$
in the sense of
\begin{align}
\mc F^* &= \Theta^* \mc G_* \Theta.
\end{align}
\end{proposition}
\begin{proof}
\begin{align}
\Theta^* \mc G_* \Theta
&= \Theta^* \mc E_{\tilde\sigma}^{-1} \mc G \mc E_{\tilde\sigma}\Theta 
&
(\textrm{by definition of $\mc G_*$})
\\
&= \mc E_{\sigma}^{-1} \Theta^* \mc G \Theta \mc E_{\sigma}
&
(\textrm{by Eq.~(\ref{steady2}), Lemma~\ref{lem_petz},
$\Theta^* = \Theta^{-1}$})
\\
&= \mc E_{\sigma}^{-1} (\Theta^* \mc G^* \Theta)^* \mc E_{\sigma}
&
((\mc A\mc B)^*  = \mc B^* \mc A^*)
\\
&= \mc E_{\sigma}^{-1} (\mc F_*)^* \mc E_{\sigma}
&
(\textrm{by Eq.~(\ref{reverse_map})})
\\
&= \mc E_{\sigma}^{-1} (\mc E_\sigma^{-1} \mc F \mc E_\sigma)^* \mc E_{\sigma}
&
(\textrm{by definition of $\mc F_*$})
\\
&= \mc E_{\sigma}^{-1} \mc E_\sigma \mc F^* \mc E_\sigma^{-1} \mc E_{\sigma}
&
((\mc A\mc B)^*  = \mc B^* \mc A^*, \mc E_\sigma = \mc E_\sigma^*)
\\
&= \mc F^*.
\end{align}
\end{proof}

\bibliography{onsager9_QST}



\end{document}